\documentclass[12pt]{article} 
\usepackage{graphicx,amsmath, amssymb, verbatim}
\newtheorem{proposition}{Proposition}

\newtheorem{definition}{Definition}
\newtheorem{example}{Example}
\newtheorem{conjecture}{Conjecture}

\newcommand {\C } {\mathbb{C}} 
\newcommand{\qed}{\hfill $\Box$ \hfill \\}
\newcommand {\p } {\mathbb{P}} 

\newcommand {\R } {\mathbb{R}} 

\newcommand {\can} {K_{\p^2}} 
\newcommand {\loc} {$\mathcal O(-1)\oplus \mathcal O(-1)\rightarrow 
\p^1$}
\newcommand {\Z} {\mathbb{Z}} 
\DeclareMathOperator{\Res}{\mathrm{Res}}

\newcommand{\ba}{\begin{eqnarray}}
\newcommand{\ea}{\end{eqnarray}}
\newcommand{\no}{\nonumber}

\def\d{{\partial}}

\begin{document} 

\title{\bf{Extending the Picard-Fuchs system of local mirror symmetry 
}}

\author{Brian Forbes,\; Masao Jinzenji \\ \\ \it Division of 
Mathematics, Graduate School of Science \\ \it Hokkaido University \\ 
\it 
Kita-ku, Sapporo, 060-0810, Japan \\ \it brian@math.sci.hokudai.ac.jp\\
\it jin@math.sci.hokudai.ac.jp }

\date{March, 2005} 

\maketitle 

\begin {abstract} 
We propose an extended set of differential operators for local mirror 
symmetry. If $X$ is Calabi-Yau such that $\dim H_4(X,\Z)=0$, then we 
show that our operators fully describe mirror symmetry. In the process, 
a 
conjecture for intersection theory for such $X$ is uncovered. We also 
find new operators on several examples of type $X=K_S$ through similar 
techniques. In addition,  open string PF systems are considered. 

\end {abstract} 

\section{Introduction.}

For some time now, mirror symmetry has been successfully used to make 
enumerative predictions on certain Calabi-Yau manifolds. While mirror 
symmetry for compact Calabi-Yau's has been extensively studied, local 
mirror symmetry is relatively new, and a complete formulation does not 
yet 
exist. 

The first unified treatment of local mirror symmetry was written down 
in \cite{CKYZ}, in the case that the space looks like $X=K_S$, where 
$S$ 
is a Fano surface and $K_S$ is its canonical bundle. Very recently 
\cite{H}, the work of \cite{CKYZ} was formulated more mathematically. 
With 
the ideas of \cite{H}, we are able to determine all information 
relevant for mirror symmetry directly from the Picard-Fuchs equations 
of the 
mirror. These techniques are limited to the case that $X$ satisfies 
$b_2(X)=b_4(X)$.

The aim of this paper is to further the program of local mirror 
symmetry. We propose a new set of differential operators, whose 
solutions 
contain the usual local mirror symmetry solutions as a subset. In the 
event 
that the space in question contains no 4 cycle, the new operators 
completely solve the problem for an arbitrary number of K\"ahler 
parameters. 
For the more traditional mirror symmetry constructions of \cite{CKYZ}, 
our methods still complete missing data; however, a general formulation 
here is lacking. 

The key point in the construction of the extended Picard-Fuchs 
system is the determination of triple intersection numbers of K\"ahler 
classes for open Calabi-Yau manifolds. Up to now, a natural definition 
of 
triple intersection numbers of K\"ahler classes on open Calabi-Yau 
manifolds 
is not known, but in this paper, we search for triple intersection 
numbers that are 
``natural'' from the point of view of mirror symmetry in the sense of 
the following conjecture:
\begin{conjecture}
Consider the A-model on a Calabi-Yau threefold $X$. 
Let $u_{i}$ be the logarithm of the B-model complex deformation 
parameter 
$z_{i}$ obtained from the toric construction of the mirror Calabi-Yau 
manifold 
$\hat{X}$. 
Then the  B-model Yukawa coupling $C_{u_{i}u_{j}u_{k}}$ 
of $\hat{X}$ obtained from the A-model Yukawa coupling of $X$ 
is a rational function in $z_{i}=\exp(u_{i})$.
Moreover, its denominator includes the divisor of the defining equation 
of the 
discriminant locus of $\hat{X}$.    
\end{conjecture}  
Since the triple intersection number is just the constant term of the  
B-model Yukawa coupling $C_{u^{i}u^{j}u^{k}}$, the above conjecture 
imposes 
constraints on these triple intersection numbers. 
In this paper, we regard these triple intersection numbers as 
intersection 
numbers of some minimal compactification $\overline{X}$ of $X$. 
With these 
intersection 
numbers in hand, we can construct a quantum cohomology ring of 
$\overline{X}$ 
and an
associated Gauss-Manin system. In the examples treated in this paper, 
this 
quantum cohomology ring satisfies Poincare duality as a compact 3-fold, 
even if $X$ is an open Calabi-Yau 3-fold. With this Gauss-Manin system, 
we can write down differential equations for $\psi_{0}(t^{i})$, the 
function 
associated to the identity element of the quantum cohomology ring. Then 
we can rewrite these differential equations by using the mirror map 
$t_{i}=t_{i}(z_{*})$. Our assertion in this paper is that the 
differential 
equations so obtained are the extension of the Picard-Fuchs operators 
obtained from the standard toric construction of $\hat{X}$. Moreover, 
the 
extended Picard-Fuchs system has all the properties that the 
Picard-Fuchs 
system
associated to a compact Calabi-Yau 3-fold should have: a 
unique triple log solution, Yukawa couplings, etc. 

One 
problem 
in the construction is that in some cases, the constraints obtained 
from 
the above conjecture are not strong enough to determine all the triple 
intersection numbers. In other words, we have some real moduli 
parameters in the triple 
intersection numbers. Yet, we can still construct the extended 
Picard-Fuchs
system for each value of the moduli parameters, and these systems have 
all the properties desired for a Picard-Fuchs system associated to a 
compact Calabi-Yau 3-fold. In the case that $b_{4}(X)=b_{6}(X)=0$, we 
find 
unique triple intersection numbers compatible with the above conjecture 
by 
considering the change of the prepotential under flops. Hence, our 
construction 
of an extended Picard-Fuchs system has no ambiguity in this situation.   

Here is the organization of the paper. Section 2 spells out the 
generalities of the Gauss-Manin system and intersection theory for open 
Calabi-Yau manifolds. In Section 3, we thoroughly 
consider $\mathcal O(-1)\oplus \mathcal O(-1)\rightarrow \p^1$, giving 
a PF 
operator for mirror symmetry and a geometric view of the meaning of the 
operator. Section 4 is the generalization, which spells out a 
conjecture on how to deal with all $X$ such that $\dim H_4(X,\Z)=0$. 
This is subsequently applied to several cases and shown to produce the 
expected 
results. Section 5 explores the application of our techniques to open 
string 
theory on $\mathcal O(-1)\oplus \mathcal O(-1)\rightarrow \p^1$, while 
Sections 6 and 7 work through examples of type $K_S$. Some of the 
results for our more unwieldy examples are collected in the appendices.

\bigskip

$\bf{Acknowledgements.}$ 
We first would like to thank Fumitaka Yanagisawa for giving us a lot of 
help 
on computer programming. We would also like to thank Professor Martin Guest for discussions on quantum cohomology. 
 B.F. would like to thank Professor Shinobu Hosono for helpful conversations, and 
Martijn van Manen for computer assistance. The research of B.F. was 
funded by COE grant of Hokkaido University. The research of M.J. is 
partially 
supported by JSPS grant No. 16740216.

\bigskip

\section{ The Main Strategy: Overview of the Gauss-Manin System.}
Suppose that we have obtained ``natural'' classical triple intersection 
numbers 
$\int_{\overline{X}}k_{a}\wedge k_{b}\wedge k_{c}$ for an open 
Calabi-Yau 
3-fold $X$ under the assumption of Conjecture 1, and that we know the 
instanton part of the prepotential for $X$. Let us denote this 
instanton part by $\mathcal 
F^{inst}(t_{*})$ , where $t_{a}$ is the K\"ahler deformation 
parameter associated to the  K\"ahler form $k_{a} \;(a=1,\cdots, 
h^{1,1}(X))$. 
With this data, we can construct an A-model Yukawa coupling for $X$:
\begin{equation}
Y_{abc}(t_{*})= \int_{\overline{X}}k_{a}\wedge k_{b}\wedge k_{c}
+\frac{{\d}^{3}\mathcal F^{inst}(t_{*})}{\d t_{a}\d 
t_{b}\d t_{c}}.
\end{equation}
Using the classical intersection numbers 
$\int_{\overline{X}}k_{a}\wedge k_{b}\wedge k_{c}$, 
we can  
construct a basis $m_{\alpha}\;(\alpha=1,\cdots ,h^{1,1}(X)) $ of 
$H^{4}(\overline{X},\Z)$ that has the
following property:
\begin{equation}
\eta_{a\alpha}:=\int_{\overline{X}}k_{a}\wedge 
m_{\alpha}=\delta_{a\alpha}.
\end{equation}
With this setup, we obtain a (virtually compact) quantum cohomology 
ring
of $\overline{X}$, 
\begin{eqnarray}
k_{a}\circ 1&=&k_{a},\no\\
k_{a}\circ 
k_{b}&=&\sum_{c,\gamma}Y_{abc}(t_{*})\eta^{c\gamma}m_{\gamma}=
\sum_{\gamma}Y_{ab\gamma}(t_{*})m_{\gamma},\no\\
k_{a}\circ m_{\alpha}&=&Y_{a\alpha 0}v=\delta_{a\alpha}v,\no\\
k_{a}\circ v&=&0,
\label{qcv}   
\end{eqnarray}
where we used standard property of quantum cohomology ring:
\begin{equation}
Y_{a\alpha0}=\eta_{a\alpha}.
\end{equation}
Here $v$ is the volume form of $\overline{X}$ and  
we use the subscript $0$ to denote the identity $1$ of 
$H^{*}(\overline{X},\Z)$.
Then we consider the 
associated Gauss-Manin system:
\begin{eqnarray}
\d_{a}\psi_{0}&=&\psi_{a},\no\\
\d_{a}\psi_{b}&=&\sum_{c,\gamma}Y_{abc}(t_{*})\eta^{c\gamma}\psi_{\gamma}=
\sum_{\gamma}Y_{ab\gamma}(t_{*})\psi_{\gamma},\no\\
\d_{a}\psi_{\alpha}&=&Y_{a\alpha 
0}\psi_{v}=\delta_{a\alpha}\psi_{v},\no\\
\d_{a}\psi_{v}&=&0.
\label{qcgm}   
\end{eqnarray}
Next, we consider the inverse matrix $(Y^{-1}_{a}(t_{*}))^{bc}$ of 
$(Y_{a}(t_{*}))_{bc}:=Y_{abc}(t_{*})$. From (\ref{qcgm}), we obtain,
\begin{equation}
\psi_{\alpha}=\sum_{b}(Y^{-1}_{a}(t_{*}))^{\alpha 
b}\d_{a}\d_{b}\psi_{0}.
\label{0}
\end{equation}
Since, $\psi_{\alpha}$ is unique for each ${\alpha}$, we have to 
impose integrability conditions:
\begin{equation}
\sum_{c}(Y^{-1}_{a}(t_{*}))^{\alpha c}\d_{a}\d_{c}\psi_{0}
=\sum_{c}(Y^{-1}_{b}(t_{*}))^{\alpha c}\d_{b}\d_{c}\psi_{0}\;\;(a\neq 
b)
\label{1}
\end{equation} 
for any $a,\; b\in\{1,2,\cdots,h^{1,1}(X)\}$.
We have another integrability condition from the third line of 
(\ref{qcgm}): 
\begin{eqnarray}
&&\d_{a}
\bigr(\sum_{c}(Y^{-1}_{a}(t_{*}))^{ac}\d_{a}\d_{c}\psi_{0}\bigr)
=\d_{b}
\bigr(\sum_{c}(Y^{-1}_{b}(t_{*}))^{b c}\d_{b}\d_{c}\psi_{0}\bigr),
\;\;(a\neq b)\no\\
&&\d_{b}
\bigr(\sum_{c}(Y^{-1}_{a}(t_{*}))^{ac}\d_{a}\d_{c}\psi_{0}\bigr)=0,
\;\;(a\neq b).
\label{2}
\end{eqnarray} 
for any $a,\; b\in\{1,2,\cdots,h^{1,1}(X)\}$.
Finally, we can derive differential equations from the fourth line of 
(\ref{qcgm}):
\begin{equation}
\d_{a}^2
\bigr(\sum_{c}(Y^{-1}_{a}(t_{*}))^{ac}\d_{a}\d_{c}\psi_{0}\bigr)
=0.
\label{3}
\end{equation}
Our strategy in this paper is to translate the equations (\ref{1}), 
(\ref{2}) 
and (\ref{3}) in terms of the mirror map 
\begin{equation}
t_{a}=t_{a}(z_{*})
\end{equation}
into the differential equations of the complex deformation parameters 
$z_{a}$ 
of the mirror manifold $\hat{X}$. Of course, in the one parameter case, 
the 
equations
(\ref{1}) and (\ref{2}) become trivial, and the only nontrivial 
equation 
is
\begin{equation}
\label{afunexample}
\d^{2}_{t}\biggl(\frac{1}{Y_{ttt}}\biggr)\d^{2}_{t}\psi_{0}(t)=0,
\end{equation}
which is well-known from the literature. 

In the following, we will explicitly compute 
(\ref{1}), (\ref{2}) and (\ref{3}) in many examples, and we find that 
these equations are highly degenerate. Therefore, in this paper, 
we will choose the minimal independent set of equations for an extended 
Picard-Fuchs 
system.  

By construction, the  Picard-Fuchs system so obtained  has a solution 
space 
given by 
\begin{equation}
\Big(1,t_1,\dots,t_{h^{1,1}(X)},\frac{\d \mathcal F}{\d 
t_1},\dots,\frac{\d \mathcal F}{\partial t_{h^{1,1}(X)}},2\mathcal 
F-\sum_{a=1}^{h^{1,1}(X)}t_{a}
\frac{\d \mathcal F}{\d t_a}\Big).
\end{equation}
And this is exactly the data one would like from mirror symmetry.

\section{Mirror symmetry for local $\p^1$.}

\subsection{A Picard-Fuchs operator for local $\p^1$.}

Before diving into the details of Gauss-Manin systems and the like, we 
will first take a simple-minded look at a familiar example, namely 
$\mathcal O(-1)\oplus 
\mathcal O(-1)\rightarrow \p^1$. We will see that in trying to apply 
the techniques of local mirror symmetry to this basic case, we are 
inevitably led to introduce the generalized intersection theory 
explained in 
the introduction. In fact, this is the example that originally 
motivated the investigations of this paper.

Recall the symplectic quotient definition of $\mathcal O(-1)\oplus 
\mathcal O(-1)\rightarrow \p^1$:
\begin{equation}
\label{localp1}
X= \{(w_1,\dots,w_4)\in \C^4-Z : 
|w_1|^2+|w_2|^2-|w_3|^2-|w_4|^2=r\}/S^1.
\end{equation}
Above, $Z=\{w_1=w_2=0\}$, $r \in \R^+$ and 
$$
S^1:(w_1,\dots,w_4)\rightarrow 
(e^{i\theta}w_1,e^{i\theta}w_2,e^{-i\theta}w_3,e^{-i\theta}w_4).
$$

We can naively employ the methods of \cite{H} to produce a Picard-Fuchs 
operator associated to the mirror Calabi-Yau $\hat X$ of $X$. The 
family $\hat X$ 
is described as \cite{HV}
\begin{equation}
\label{mirror}
\hat X_z=\{(u,v,y_1,y_2)\in \C^2\times (\C^*)^2: 
uv+1+y_1+y_2+zy_1y_2^{-1}=0\}.
\end{equation}
Then \cite{H} provides a recipe for dealing with non-compact period 
integrals for such an $\hat X$. They are defined by 
$$
\Pi_{\Gamma}(z)=\int_{\Gamma}\frac{dudvdy_1dy_2/(y_1y_2)}{uv+1+y_1+y_2+zy_1 
y_2^{-1}}
$$
for $\Gamma \in H_4(\C^2\times (\C^*)^2-\hat X,\Z)$. As usual, we 
utilize 
the GKZ formalism in order to exhibit an differential operator which 
annihilates the $\Pi_{\Gamma}$. One finds 
\begin{equation}
\label{localPF}
\mathcal D = (1-z)\theta^2, \ \ \ \theta=z\frac{d}{dz}
\end{equation}
as the relevant PF operator.

This is a puzzling situation. Clearly, the solutions of $\mathcal D 
f=0$ are given by $\{1, \log z\}$. This is sensible, because noncompact 
PF 
systems always have a constant solution \cite{CKYZ}, and the mirror map 
is trivial in this case, leading to a $\log z$ solution. However, there 
is no double logarithmic solution, because $\mathcal D$ is only of 
order 2. Hence, we have no function $\mathcal F$ with which to count 
holomorphic curves on $X$! But, since $X$ contains exactly 1 
holomorphic 
curve, we know that the sought after function should be of the form 
\begin{equation}
\label{prepotential}
\mathcal F(z)= K\frac{(\log z)^3}{6}+ \sum_{n >0}\frac{z^n}{n^3}.
\end{equation}
Here $K$ is a classical triple intersection number for 
$\p^1\hookrightarrow X$.

Also, notice that the leading factor of $1-z$ in front of $\mathcal D$, 
while naturally appearing through the techniques of the GKZ formalism, 
is auxiliary to the solution set of $\mathcal D$. 

At this point, we can gain a bit of insight from the compact 
case. Recall \cite{CK} that in the event of a compact Calabi-Yau $X$ 
with one K\"ahler parameter, there is always a flat coordinate $t$ in 
which the Picard-Fuchs operator for the mirror family is given as 
\begin{equation}
\label{compactPF}
\mathcal D_{compact}(t)= \partial_t^2\Big(\frac{1}{Y}\Big)\partial_t^2,
\end{equation}
which is the same as the formula (\ref{afunexample}).
This is reminiscent of our situation (\ref{localPF}), upon making the 
identification $t=\log z$.

If one surrendered to the impulse of emulating the above compact 
expression, one would be compelled to work with the following modified 
differential operator:
$$
\mathcal D \longrightarrow \mathcal D'= \theta^2\mathcal D.
$$
Rewrite this as
$$
\theta^2(1-z)\theta^2 = \theta^2\Big(\frac{1}{1-z}\Big)^{-1}\theta^2.
$$
By comparison with (\ref{compactPF}), it is natural to identify the 
Yukawa coupling $Y$ of $\mathcal O(-1)\oplus \mathcal O(-1)\rightarrow 
\p^1$ as
\begin{equation}
\label{yukawa}
Y=\frac{1}{1-z}.
\end{equation}
And indeed, the condition that $Y=\theta^3\mathcal F$, which follows 
from the form of $\mathcal D'$, yields the expected function $\mathcal 
F$ 
(\ref{prepotential}). Then the resultant period vector is
$$
\Pi'=(1,\log z,\theta \mathcal F, 2\mathcal F - (\log z)\theta \mathcal 
F),
$$
which is the period vector encountered when dealing with compact 
Calabi-Yaus. Hence, we have found a `cure' for mirror symmetry on 
$X=\mathcal 
O(-1)\oplus \mathcal O(-1) \rightarrow \p^1$; the new operator 
$\theta^2 \mathcal D$ reproduces all relevant data to describe mirror 
symmetry 
for $X$. 

We can also view this $\theta^2\mathcal D$ from the vantage of the 
Frobenius method. The geometry of $X$ is determined by the set of 
vertices 
$\{\nu_0,\dots,\nu_3\}=\{(0,0),(1,0),(0,1),(1,1)\}$, together with a 
choice of triangulation of the resulting toric graph. These give rise 
to 
the lattice vector $l=(1,1,-1,-1)$, and identify $z$ as the correct 
variable on the complex moduli space of $\hat X$. 

Then the solutions of our extended PF operator $\theta^2\mathcal D$ can 
be generated, via the Frobenius method, from the function
$\omega_0(z,\rho)=\sum_{n\ge0}c(n,\rho)z^{n+\rho}$, with
$$
c(n,\rho)=(\Gamma(1+n+\rho)^2\Gamma(1-n-\rho)^2)^{-1}.
$$
It is a simple matter to verify that
$$
\Pi' =(\omega_0(z,0), \ \partial_{\rho} \omega_0(z,\rho)|_{\rho=0}, \ 
\partial_{\rho}^2 \omega_0(z,\rho)|_{\rho=0}, \ \partial_{\rho}^3 
\omega_0(z,\rho)|_{\rho=0}).
$$
Clearly this had to be the case, since the extension of the original 
$\mathcal D$ on the left by each factor of $\theta$ adds one more 
Frobenius-generated solution.

There is, however, one additional subtlety here; the constant $K$ from 
eqn. (\ref{prepotential}) was not determined. It is expected from 
physics that such a $K$ ought to be ambiguous. However, there is a 
unique 
choice which is compatible with the solutions of $\theta^2\mathcal D$, 
which turns out to be $K=1$. This is the same as was used in the 
physical 
considerations of \cite{V}. Later, we will see that the choice more 
natural for generalization is $K=1/2$.

Next, we will take up the question of the geometric meaning of the 
solutions of this new operator.

\subsection{PF extensions and Riemann surfaces.}

In physics literature \cite{AV} \cite{H}, a frequently used technique 
of local mirror symmetry is to consider periods on a Riemann surface 
$\Sigma \hookrightarrow \hat X$, rather than periods of the full mirror 
geometry $\hat X$. In this section, we will review evidence in favor of 
this 
approach.

Looking back at the mirror geometry (\ref{mirror}), this can be 
rewritten as 
$$
\hat X_z=\{ uv+1+y_1+y_2+zy_1y_2^{-1}= uv +f(z,y_1,y_2)=0\},
$$
which is a hypersurface in  $\C^2\times (\C^*)^2$. Notice that there is 
an imbedded Riemann surface in this space, defined as
\begin{equation}
\label{riemann}
\Sigma_z=\{(y_1,y_2)\in(\C^*)^2: f(z,y_1,y_2)=0\}.
\end{equation}
In fact, this statement applies not only to the local $\p^1$ case, but 
to all toric local mirror symmetry constructions \cite{AV}. The only 
difference is that there may be more complex moduli involved; in such 
cases, we can simply set $z=(z_1,\dots,z_n)$, where now each $z_i$ is a 
complex structure modulus.

Now take $(a_0,\dots,a_3)$ as homogeneous coordinates on the moduli 
spaces of $\hat X$ and $\Sigma$, i.e.  
$$
\hat X_a=\{ uv+a_0+a_1y_1+a_2y_2+a_3zy_1y_2^{-1}= uv +f(a,y_1,y_2)=0\}.
$$
Recall that the GKZ operators are differential operators $\{\mathcal 
L_i\}$ in the variables $a$, such that
$$
\mathcal L_i\int_{\Gamma}\frac{dudvdy_1dy_2/(y_1y_2)}{uv 
+f(a,y_1,y_2)}=0, \ \ \forall i.
$$
We can recover the PF operators from the GKZ operators via a canonical 
reduction on the homogeneous moduli space.

With these things in mind, we note the general
\begin{proposition}
\label{reduction}
The GKZ operators associated to the geometry $\hat X$ are the same as 
those 
associated to $\Sigma$.
\end{proposition}
\it Proof. \rm Notice that $\Sigma$ is a complex dimension 1 noncompact 
Calabi-Yau manifold. In particular, it makes sense to define the period 
integrals of $\Sigma$ as
$$
\Pi_{\gamma}^{\Sigma}(a)=\int_{\gamma}\frac{dy_1dy_2/(y_1y_2)}{f(a,y_1,y_2)}
$$
with $\gamma \in H_2((\C^*)^2-\Sigma,\Z)$. Let $\mathcal L$ be a GKZ 
operator on the moduli space of $\Sigma_a$, so that 
$$\mathcal L \Pi_{\gamma}^{\Sigma}(a)=0.$$
Recall that the period integrals of $\hat X_a$ are given as 
\begin{equation}
\label{yourmom}
\Pi_{\Gamma}^{\hat 
X}(a)=\int_{\Gamma}\frac{dudvdy_1dy_2/(y_1y_2)}{uv+f(a,y_1,y_2)}
\end{equation}
for $\Gamma \in H_4(\C^2\times (\C^*)^2-\hat X,\Z)$. Then it is clear 
that 
we must also have
$$
\mathcal L \Pi_{\Gamma}^{\hat X}(a)=0,
$$
because the additive factor of $uv$ in the period integrals of $\hat X$ 
is 
independent of $a$. Clearly the converse of this statement is also 
true, 
so the proposition follows.
\qed

Note that, as pointed out in \cite{H}, there us a subtlety in terms of 
the scaling properties of the period integrals on $\hat X_a$ and 
$\Sigma_a$. 
Further, this scaling difference implies that the PF operators we 
derive from the above $\mathcal L$ will be different on $\Sigma_z$ and 
$\hat X_z$. In the following, we will ignore this point, and carry on 
as 
though the period integrals on $\Sigma_z$ actually reproduce the same 
PF 
operators. 

As the geometry of $\Sigma$ is far simpler than that of $\hat X$, this 
proposition will greatly aid the search for a geometric interpretation 
of 
the formal procedure $\mathcal D \rightarrow \theta^2 \mathcal D$ on 
Picard-Fuchs operators. We will explore this in the next section.

\subsection{Geometric interpretation through the Riemann surface.}

First, we will give a brief description of what ``adding extra period 
integrals" means (which we are doing, by raising the power of the PF 
operator) in the context of the space $\hat X$. This follows the lead 
of e.g. 
\cite{HV},\cite{CIV}.

Recall that mirror symmetry between the spaces $X$ and $\hat X$ means, 
in 
particular, that 
$$
\dim H^{1,1}(X)=\dim H^{2,1}(\hat X).
$$
Hence, for every 2 cycle of $X$, we can expect a mirror 3 cycle of 
$\hat X$. 
Let $\dim H_3(\hat X,\Z)=n$, and take $\Gamma_i,  \Gamma_j \in H_3(\hat 
X,\Z)$ 
with Poincar\'e duals $\alpha_i, \alpha_j \in H^3(\hat X,\Z)$. Then 
there is a 
symplectic structure on $H_3(\hat X,\Z)$, defined by the intersection 
pairing
$$
(\Gamma_i,\Gamma_j)=\int_{\hat X} \alpha_i \wedge \alpha_j.
$$

In the compact case, we can find a basis 
$\{\Phi_1,\dots,\Phi_{n/2},\Psi_1,\dots,\Psi_{n/2}\}$ for $H_3(\hat 
X,\Z)$ satisfying
$$
(\Phi_i,\Phi_j)=(\Psi_i,\Psi_j)=0 \ , \ \ \  
(\Phi_i,\Psi_j)=\delta_{ij}.
$$

However, there is no such nice construction for the noncompact case. In 
fact, we can explicitly exhibit this failure in the example we are 
considering, $\mathcal O(-1)\oplus \mathcal O(-1)\rightarrow \p^1$. 
First, 
rewrite the equation for the mirror $\hat X$:
$$
\hat X_z= \{\tilde u  v + z + \tilde y_1 + \tilde y_2 + \tilde  y_1 
\tilde 
y_2 =0\}
$$
where we have taken $\tilde y_i = y_i^{-1}$ and $\tilde u = u/ y_1  
y_2$. Set $z=1-a$, where $a\in \R^+.$ Then 
$$
\hat X_z= \{\tilde u  v + 1 + \tilde y_1 + \tilde y_2 + \tilde  y_1 
\tilde 
y_2 =a\},
$$
and we can identify a 3 cycle 
$$ 
\Gamma_1= \hat X_z\cap\{\tilde u= \bar v, \tilde y_2= 
\bar{\tilde{y_1}}\}= 
\{v\bar v + (1+ \tilde y_1)\overline{(1+ \tilde y_1)}=a\}.
$$
It is easy to verify that this cycle has no symplectic dual in 
$H_3(\hat X,\Z)$.

Of course, there is a noncompact symplectic dual for $\Gamma_1$. Let $p 
\in \Gamma_1$, and take
$\tilde \Gamma_1 = (N_{\Gamma_1 / \hat X})_p$. Then $\tilde \Gamma_1$ 
intersects $\Gamma_1$ in a point, and could be thought of as a dual; 
however, 
integrals over $\tilde \Gamma_1$ may not be well defined. Instead take
$$
\tilde \Gamma_1^{\lambda}=\{v\in (N_{\Gamma_1 / {\hat X}})_p: |v| \le 
\lambda\}, \ \ \lambda \in \R^+.
$$
Set
$$
\Omega_{\hat 
X}=\Res_{\{uv+f(z,y_1,y_2)=0\}}\Big(\frac{dudvdy_1dy_2/(y_1y_2)}{uv+f(z,y_1,y_2)}\Big).
$$
Then the proposal of \cite{HV} is that we should also consider 
integrals of the form
$$
\int_{\tilde \Gamma_1^{\lambda}} \Omega_{\hat X}
$$
as periods of the noncompact geometry $\hat X$. Mathematically, this 
means 
that the definition of noncompact period integrals of $\hat X$ are to 
be 
taken as all $\int_{\Gamma}\Omega_{\hat X}$ for $\Gamma \in H_3(\hat 
X,\Z)\oplus 
(H_3(\hat X,\Z))_c$. Here, the subscript $c$ indicates compactly 
supported 
homology.

In view of Proposition 1, one can make the following
\begin{definition}
\label{riemannperiods}
Let $\hat X$ be the noncompact Calabi-Yau hypersurface
$$
\hat X=\{(u,v,y_1,y_2)\in \C^2\times (\C^*)^2: uv+f(z,y_1,y_2)=0\}
$$ 
and $\Sigma$ the imbedded Riemann surface $\Sigma_z={\hat X}\cap 
\{u=v=0\}.$
Then the \rm period integrals of \it $\hat X$ are defined to be
$$
\Pi_{\gamma}(z)= 
\int_{\gamma}\Res_{f=0}\Big(\frac{dy_1dy_2/(y_1y_2)}{f(z,y_1,y_2)}\Big)
$$
for $\gamma \in H_1(\Sigma, \Z) \oplus (H_1(\Sigma, \Z))_c$.
\end{definition}

In the next subsection, we will apply this definition to $\mathcal 
O(-1)\oplus \mathcal O(-1)\rightarrow \p^1$, and see that the explicit 
evaluation of the period integrals on $\Sigma$ gives the same answer as 
the 
extended PF operator of section 3.1.

\subsection{Period integrals for local $\p^1$.}

Before describing the cycles of $\Sigma$ and computing their associated 
integrals, we will need to make use of the following proposition. This 
will give a $(1,0)$ form $\alpha \in H_1(\log (\Sigma),\Z)$, which can 
be integrated over lines in $\Sigma$. Note that, as was the case of the 
previous proposition, the validity is for all Riemann surfaces 
appearing as mirrors of the toric local mirror symmetry construction.

Some of the arguments below can be found in \cite{KKV}.
\begin{proposition}
\label{sigmaintegral}
Let $\Sigma$ be as above, and choose $\gamma \in H_1(\Sigma, \Z) \oplus 
(H_1(\Sigma, \Z))_c$.
Then 
$$
\int_{\gamma}\Res_{f=0}\Big(\frac{dy_1dy_2/(y_1y_2)}{f(z,y_1,y_2)}\Big)= 
-\int_{\gamma} \log y_2 \frac{dy_1}{y_1}=-\int_{\gamma} \log y_1 
\frac{dy_2}{y_2}.
$$
\end{proposition}
\it Proof. \rm Let $T(\gamma)$ be a tubular neighborhood of $\gamma$ in 
$\Sigma$. One easily sees that the PF operators of 
$$
\int_{T(\gamma)}\frac{dy_1dy_2/(y_1y_2)}{f(z,y_1,y_2)} \ \ \ \rm and 
\it  \ \ \ \int_{T(\gamma)} \log (f) dy_1dy_2/(y_1y_2)
$$ 
are in fact the same. Then, if we assume that Definition 1 gives an 
equivalence between PF solutions and period integrals, we have
$$\int_{\gamma}\Res_{f=0}\Big(\frac{dy_1dy_2/(y_1y_2)}{f(z,y_1,y_2)}\Big)=\int_{\gamma}\Res_{f=0}\Big( 
\log (f)dy_1dy_2/(y_1y_2)\Big)=
$$
$$
-\int_{\gamma}\Res_{f=0}\Big(d(\log (f)) \log y_2 
\frac{dy_1}{y_1}\Big)=-\int_{\gamma}\Res_{f=0}\Big( \frac{df}{f} \log 
y_2 
\frac{dy_1}{y_1}\Big)=
$$
$$
-\int_{\gamma} \log y_2 \frac{dy_1}{y_1}.
$$
From the first to the second line, we have integrated by parts, and 
then the residue around $f=0$ was taken from the second to third line. 

This argument applies equally well upon exchanging $y_1$ and $y_2$.
\qed

With this proposition in hand, we can take up the task of working out 
period integrals on local $\p^1$. Recall that the original, unmodified 
PF operator for this space was found to be
$$
\mathcal D = (1-z)\theta^2,
$$
with solution set $\{1,\log z\}$. We will at first content ourselves 
with finding cycles $\gamma_0,\gamma_1 \in H_1(\Sigma, \Z)$ whose 
period 
integrals reproduce these solutions.

To this end, note that the defining equation
\begin{equation}
\label{riemannsurface}
\Sigma = \{(y_1,y_2)\in (\C^*)^2: 1+y_1+y_2+zy_1y_2^{-1}=0\}
\end{equation}
can be solved in 2 ways:
$$
y_1=\frac{-1-y_2}{1+zy_2^{-1}}, \ \ \ y_2^{\pm}=\frac{-1-y_1 \pm 
\sqrt{(1+y_1)^2-4zy_1}}{2}.
$$
Now, since $y_1$ and $y_2$ are $\C^*$ variables, we can define three 
homology elements from these equations:
$$
\gamma_0=\Big\{(y_1,y_2)\in (\C^*)^2: y_1=\frac{-1-y_2}{1+zy_2^{-1}} \ 
, \ \ |y_2|=\epsilon\Big\},
$$
$$
\tau_{\pm}=\Big\{(y_1,y_2)\in (\C^*)^2: y_2^{\pm}=\frac{-1-y_1 \pm 
\sqrt{(1+y_1)^2-4zy_1}}{2} \ , \ \ |y_1|=\epsilon\Big\}.
$$
Then, two of these must be responsible for the solution set $\{1,\log 
z\}$. To motivate the correct choice of cycles, let us first look 
closely at the mirror construction that originally provided equation 
(\ref{riemannsurface}).

Starting with the description $\mathcal O(-1)\oplus \mathcal 
O(-1)\rightarrow \p^1=$
$$
\{(w_1,\dots,w_4)\in \C^4-Z : |w_1|^2+|w_2|^2-|w_3|^2-|w_4|^2=r\}/S^1,
$$
from \cite{AV}, the mirror geometry can be characterized as 
$$
\{uv+x_1+\dots+x_4=0: x_1x_2x_3^{-1}x_4^{-1}=z, \ x_i=1 \ \rm for \ 
some \ \it i\}.
$$
Here $u,v\in \C$ and $x_i\in \C^*$. Also, the $x_i$ obey 
$|x_i|=\exp(-|w_i|^2)$, and $|z|=e^{-r}.$ To arrive at the form 
(\ref{riemannsurface}), we set $x_3=1$ and solved for $x_1$ in the 
constraint 
$x_1x_2x_3^{-1}x_4^{-1}=z$. Finally, the identification $y_1=x_4, \ 
y_2=x_2$ was 
made. 

The equation for local $\p^1$ indicates that $[w_1,w_2]$ can be taken 
as homogeneous coordinates on the $\p^1$. In the locus where 
$w_3=w_4=0$, we thus have a bound $0 \le |w_2|^2 \le r$. Then 
$|y_2|=\exp(-|w_2|^2)$ implies that $1 \ge |y_2| \ge e^{-r}=|z|$; 
taking these 
considerations together, we may accurately label the following figure:
\begin{figure}[htbp]
\centering
\input{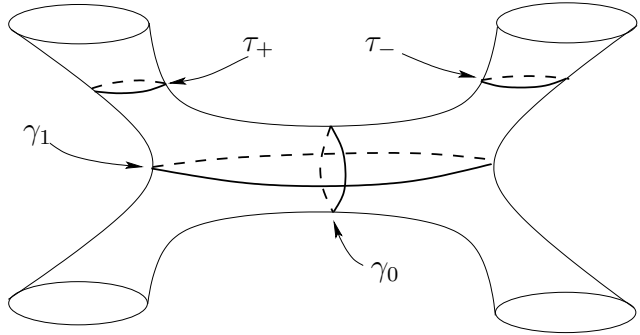}
\caption{1-cycles on $\Sigma$. $\{\gamma_0, \tau_+, \tau_-\}$ is a 
basis for $H_1(\Sigma, \Z)$. }
\label{fig:RiemannRules}
\end{figure}
\begin{proposition}
\label{solns}
Let $\gamma_0, \ \tau_{\pm}$ be as described above, and set
$$
\gamma_1= [\tau_+]+[\tau_-]
$$
with the sum taken in $H_1(\Sigma, \Z)$. Then
$$
\int_{\gamma_0} \log y_1 \frac{dy_2}{y_2}=1, \ \ \ \int_{\gamma_1} \log 
y_2 \frac{dy_1}{y_1}=\log z
$$
when appropriately normalized.

\end{proposition}
\it Proof. \rm The first integral is trivial:
$$
\int_{\gamma_0} \log y_1 \frac{dy_2}{y_2}=  \int_{|y_2|=\epsilon} \log 
\Big(\frac{-1-y_2}{1+zy_2^{-1}}\Big) \frac{dy_2}{y_2}= 
$$
$$
\int_{|y_2|=\epsilon}\Big( i\pi+ \log 
\Big(\frac{1+y_2}{1+zy_2^{-1}}\Big)\Big) \frac{dy_2}{y_2}
$$
and this is a constant. Of course, the branch cut of $\log$ must be 
taken to lie off the negative real axis. 

For the second,
$$
\int_{\gamma_1} \log y_2 \frac{dy_1}{y_1}=  \int_{\tau_+} \log y_2^+ \ 
\frac{dy_1}{y_1}+\int_{\tau_-} \log y_2^- \  \frac{dy_1}{y_1}= 
$$
$$
\int_{|y_1|=\epsilon} \log 
(y_2^+y_2^-)\frac{dy_1}{y_1}=\int_{|y_1|=\epsilon} \log 
(zy_1)\frac{dy_1}{y_1}= const+ 2\pi i\log z.
$$
\qed

Next, the existence of a new period integral, based on Definition 1, 
will be demonstrated.

\subsection{A period integral from $(H_1(\Sigma,\Z))_c.$}

Since the constraint $1 \ge |y_2| \ge |z|$ only applies in regions with 
$z_3=z_4=0$, outside of this locus, it is sensible to define a path on 
$\Sigma$ as follows. Let $\lambda < |z|$ be real, and take a smooth 
increasing function $\sigma:[0,1]:\rightarrow \Sigma$ such that 
$\sigma(0)=\lambda, \sigma(1)=z$. Then
$$
\gamma_2(\lambda)=\Big\{(y_1,y_2)\in (\C^*)^2: 
y_1=\frac{-1-y_2}{1+zy_2^{-1}}, \ \ y_2=\sigma[0,1], \ \ y_1 \ne 1 
\Big\}
$$
defines an element of $(H_1(\Sigma,\Z))_c.$

\begin{proposition}
\label{extraperiod}
Let $\Sigma,  \gamma_2$ be defined as above. Then
$$
\theta \mathcal F= \int_{\gamma_2} \log y_1 \frac{dy_2}{y_2},
$$
where $\theta = z\frac{d}{dz}$ and $\theta \mathcal F$ is the double 
logarithmic solution of the extended PF operator
$$
\mathcal D'=\theta^2(1-z)\theta^2
$$
associated to the mirror of the local  model $\mathcal O(-1)\oplus 
\mathcal O(-1)\rightarrow \p^1$.
\end{proposition}
\it Proof. \rm The computation is straightforward:
$$
\int_{\gamma_2} \log y_1 \frac{dy_2}{y_2}=\int_{\lambda}^{z} \log 
\Big(\frac{-1-y_2}{1+zy_2^{-1}}\Big)  \frac{dy_2}{y_2}=
$$
$$
\int_{\lambda}^{z} 
\Big(i\pi+\sum_{n>0}\frac{(-y_2)^n}{n}-\sum_{n>0}\frac{(-zy_2^{-1})^n}{n}\Big)\frac{dy_2}{y_2}=
$$
$$
const+(\lambda- \rm dependent) \it + \sum_{n>0}\frac{(-z)^n}{n^2}= 
\big(\theta \mathcal F\big)(-z).
$$
In order to achieve the result, we should have $z$ rather than $-z$ in 
the above. However, this is accounted for by the fact that $\arg(y_2)$ 
is not determined in local mirror symmetry, and hence we are free to 
use the variable $y_2'=e^{i\pi}y_2$ in place of $y_2$. 
\qed

Notice that, with this definition of period integrals, the logarithmic 
terms of $\theta \mathcal F$ are not uniquely determined, as they 
depend on $\lambda$. It is for this reason that $\lambda$ dependent 
terms 
are disregarded in the calculation. 

It may seem that the choice of $\gamma_2$ is artificial, since we could 
have equally well chosen an increasing function 
$\sigma':[0,1]\rightarrow \Sigma$ with 
$\sigma'(0)=1,\sigma'(1)=\lambda>1$. However, it is 
easy to show that this is equivalent; if 
$$
\gamma_2'(\lambda)=\Big\{(y_1,y_2)\in (\C^*)^2: 
y_1=\frac{-1-y_2}{1+zy_2^{-1}}, \ \ y_2=\sigma'[0,1], \ \ y_1 \ne 1 
\Big\},
$$
then
$$
\lim_{\lambda\to\infty} \ \Big(\theta \int_{\gamma_2'} \log y_1 
\frac{dy_2}{y_2}\Big)=\sum_{n>0}\frac{(-z)^n}{n},
$$
so the two approaches are interchangeable.

\subsection{Discussion.}

So far, we have considered only one example: $\mathcal O(-1)\oplus 
\mathcal O(-1)\rightarrow \p^1$. However, we have learned a great deal 
already. Firstly, the PF system cannot always be relied on to provide 
all 
the information we need for local mirror symmetry. Yet, one might be 
tempted to hope that, in general, we can recover the missing data 
through 
a simple modification of the known PF systems. Let us consider our new 
operator $\theta^2\mathcal D$ from one final perspective.

The incompleteness of the PF system on $X=\mathcal O(-1)\oplus \mathcal 
O(-1)\rightarrow \p^1$ is supposed to emerge from the noncompactness of 
the space; that is, as a result of the facts that $b_4=b_6=0$, where 
$b_i$ denotes the $i$th Betti number. On a compact space, the PF system 
will have one regular solution, $b_2$(resp.  $b_4$) logarithmic (resp. 
double logarithmic) solutions, and one ($=b_6$) triple logarithmic 
solution. On $X$, our two usual PF solutions can be summarized by a 
cohomology-valued hypergeometric series \cite{G}\cite{H0}\cite{H}
$$
\omega(z,J)=\omega_0+\omega_1(z)J.
$$
Here $\omega_0=1$, and $\omega_1=\log z$ is the mirror map. Also, $J$ 
is the cohomology element dual to $\p^1 \hookrightarrow X$. Solutions 
of 
the PF system $\mathcal Df=0$ are recovered as 
$\omega_i(z)=\frac{d^i\omega(z,J)}{dJ^i}|_{J=0}.$

 It is clear that by instead working with $\theta^2 \mathcal Df=0$, we 
have a larger cohomology-valued series as a generating function of 
solutions:
$$
\omega(z,J)=\omega_0+\omega_1(z)J+\omega_2(z)J^2+\omega_3(z)J^3.
$$
From this perspective, the addition of $\theta^2$ on the left-hand side 
of the operator $\mathcal D$ is a sort of compactification of the 
model, in that it represents the addition of a 4-cycle and a 6-cycle 
for 
$X$.

We will now generalize these ideas to situations with no 4-cycle, and 
an arbitrary number of K\"ahler classes.

\section{Mirror Symmetry for Toric Trees.}

In this section, we will show that if $X$ is a noncompact Calabi-Yau 
manifold such that $\dim H_4(X,\Z)=0,$ then we may apply our methods to 
obtain a complete system of differential equations which fully 
determines the prepotential. The Yukawa couplings take on a central 
role in the 
following.

\subsection{Ordinary Picard-Fuchs systems.}

Our first interest will be to take a look at the PF systems one would 
arrive at through use of existing local mirror symmetry techniques. We 
wish to understand exactly how much information one might recover 
through these systems alone, in order to determine an appropriate 
`fix'. 

Let us clarify what we are exploring here. Let 
$\{l^1,\dots,l^n\}\subset \Z^m$ be a choice of basis for the secondary 
fan of a noncompact 
toric Calabi-Yau threefold $X$ satisfying $\dim H_4(X,\Z)=0$. Consider 
the 
generating function
\begin{equation}
\omega(z,\rho)=\sum_{n>0}c(n,\rho)z^{n+\rho}
\end{equation}
where
\begin{equation}
c(n,\rho)^{-1}=\prod_i \Gamma\big(1+\sum_k l^k_i(n_k+\rho_k)\big). 
\end{equation}
Here we are using the convention that $l^k=(l_1^k,\dots,l_m^k).$ Then 
we want to look at the functions
\begin{equation}
\Pi_{ij}=\big(\partial_{\rho_i}\partial_{\rho_j}\omega(z,\rho)\big)|{\rho=0}.
\end{equation}
Our interest in this subsection, then, is to ascertain how much 
information we can find by looking at the $\Pi_{ij}$. In so doing, we 
will 
gain a better understanding of what to do in order to remedy mirror 
symmetry in this situation.

\begin{example} \rm

Consider the space
$$
X_1=\{-2|w_1|^2+|w_2|^2+|w_3|^2=Re(t), \ 
|w_1|^2-|w_2|^2+|w_4|^2-|w_5|^2=Re(s)\}/(S^1)^2
$$
where $(w_1,\dots,w_5)\in 
\C^5-(\{w_2=w_3=0\}\cup\{w_1=w_4=0\}\cup\{w_4=w_5=0\})$. This contains 
two curves $C_t, C_s$ with respective normal 
bundles $\mathcal O\oplus \mathcal O(-2)$ and $\mathcal O(-1)\oplus 
\mathcal O(-1)$ in $X_1$. We have that $b_2=2$ and $b_4=0$.

From the work \cite{LLLZ}, we can draw a planar trivalent graph for 
$X_1$ corresponding to the torus weights. Using the rules of that 
paper, 
the resulting graph looks like 
\begin{figure}[htbp]
\centering
\input{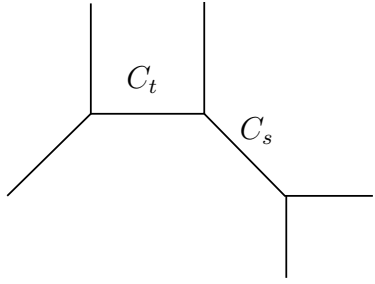}
\caption{Toric diagram for $X_1$. }
\label{aminusoneandaminustwo}
\end{figure}
Through the GKZ formalism, we have the PF operators associated to 
$X_1$:
\begin{eqnarray}
\label{PF1}
\mathcal 
D_1&=&\theta_1(\theta_1-\theta_2)-z_1(2\theta_1-\theta_2)(2\theta_1-\theta_2+1)\no\\
\mathcal 
D_2&=&(2\theta_1-\theta_2)\theta_2-z_2(\theta_1-\theta_2)\theta_2 \no\\
\mathcal D_3&=& \theta_1\theta_2-z_1z_2(2\theta_1-\theta_2)\theta_2.
\end{eqnarray}
Let $\omega(z,\rho)$ be the generating function for solutions of this, 
with logarithmic solutions $ t, s.$ 

Set $\Pi_{ij}=\partial_{\rho_i}\partial_{\rho_j}\omega|_{\rho=0}$. Then 
$\{\mathcal D_1,\mathcal D_2\}$ also has a double logarithmic solution, 
given as the linear combination
$$
W(z)=\frac{1}{2}\Pi_{11}+\Pi_{12}+\Pi_{22}.
$$

We are interested in the relationship between this double logarithmic 
solution and the prepotential $\mathcal F_1$ for $X$. From physics 
\cite{DFG}\cite{I}, the instanton part of the prepotential is
\begin{equation}
\label{myprepot}
\mathcal F_1^{inst}= 
\sum_{n>0}\frac{e^{ns}}{n^3}+\frac{e^{n(s+t)}}{n^3}-\frac{e^{nt}}{n^3}.
\end{equation}
Naturally, this is the expression gotten after use of the inverse 
mirror map. 

Let us first take a look at the $\Pi_{ij}$'s, after the insertion of 
the mirror map:
$$
\Pi_{11}(s,t)=0, 
$$
$$
\Pi_{12}(s,t)=  \sum_{n>0}\frac{e^{n(t+s)}}{n^2}-\frac{e^{ns}}{n^2},
$$
$$
\Pi_{22}(s,t)=  2\sum_{n>0}\frac{e^{ns}}{n^2}.
$$
We have neglected the logarithmic terms of each function. Notice that 
there is no linear combination of  $\Pi_{ij}$'s that we can take to 
reproduce the term $\sum_{n>0}e^{nt}/{n^2}$.

From these expressions,
$$
W(s,t)= 
\sum_{n>0}\frac{e^{n(s+t)}}{n^2}+\frac{e^{ns}}{n^2}=\frac{\partial 
\mathcal F_1^{inst}}{\partial s}.
$$

Apparently, the PF system cannot `see' the curve with normal bundle 
$\mathcal O \oplus \mathcal O(-2)$.

\end{example}

\begin{example} \rm

Next, consider the space
$$
X_2=\{|w_1|^2+|w_2|^2-|w_3|^2-|w_4|^2=Re(s_1), \ 
-|w_1|^2-|w_3|^2+|w_4|^2+|w_5|^2=Re(s_2)\}/(S^1)^2,
$$
with $(w_1,\dots,w_5)\in 
\C^5-(\{w_1=w_2=0\}\cup\{w_4=w_5=0\}\cup\{w_2=w_5=0\})$. We again have 
$b_2=2,b_4=0$, and now $\mathcal 
N_{C_i/X_2}\cong O(-1)\oplus \mathcal O(-1)$ for each $i$. Notice that 
we can flop 
from $X_2$ to $X_1$: if $l^1=(1,1,-1,-1,0)$ and $l^2=(-1,0,-1,1,1)$, 
then the combinations $l^1+l^2,-l^2$ give the secondary fan for $X_1$.
The planar trivalent toric graph for $X_2$ is given in figure 
\ref{twominusones}.
\begin{figure}[htbp]
\centering
\input{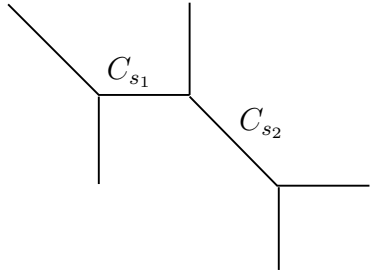}
\caption{Toric diagram for $X_2$. }
\label{twominusones}
\end{figure}
We have the PF system from the mirror manifold:
\begin{eqnarray}
\mathcal D_1&=& 
(\theta_1-\theta_2)\theta_1-z_1(-\theta_1-\theta_2)(-\theta_1+\theta_2),\no\\
\mathcal D_2&=& 
(\theta_2-\theta_1)\theta_2-z_2(-\theta_2-\theta_1)(-\theta_2+\theta_1),\no\\
\mathcal 
D_3&=&\theta_{1}\theta_{2}-z_1z_2(\theta_{1}+\theta_{2}+1)(\theta_{1}+\theta_{2}).
\end{eqnarray}
Let $s_i$ be the logarithmic solutions.
Using the same conventions as example 1, we find
$$
\Pi_{11}(s_1,s_2)=\sum_{n>0}\frac{e^{ns_1}}{n^2}-\frac{e^{ns_2}}{n^2},
$$
$$
\Pi_{12}(s_1,s_2)=0,
$$
with $\Pi_{22}=-\Pi_{11}.$ 
Again, these expressions already include the mirror map.

Let's take a look at the prepotential:
$$
\mathcal F_2^{inst}= \sum_{n\ge 
0}\frac{e^{ns_1}}{n^3}+\frac{e^{ns_2}}{n^3}-\frac{e^{n(s_1+s_2)}}{n^3}.
$$
Then we see that
$$
\Pi_{11}=\frac{\partial \mathcal F_2^{inst}}{\partial 
s_1}-\frac{\partial \mathcal F_2^{inst}}{\partial s_2}.
$$
Hence, we could recover a bit of information by using the basic 
extended system $\{\theta_1 \mathcal D_1,\mathcal D_2\}$, which has 
$\Pi_{11}$ 
as a solution. However, the cross term corresponding to $C_{s_1+s_2}$ 
cannot be detected from the $\Pi_{ij}$'s.
\end{example}

Our work with example 1 suggests the reason for the problem with the 
$C_{s_1+s_2}$ curve. To exhibit this, recall that the lattice vectors 
$\{l^1,l^2\}$ for this geometry are
$$
\begin{pmatrix}
l^1 \\ l^2 
\end{pmatrix}=
\begin{pmatrix}
	1 & 1 & -1 & -1 & 0 \\ 
	0 & -1 & -1 & 1 & 1
 \end{pmatrix}
$$
Each vector represents a curve $C_{s_i}$ in $X_2.$ Then $C_{s_1+s_2}$ 
is determined by the single vector
$$
l^1+l^2=
\begin{pmatrix}
1 & 0 & -2 & 0 & 1
\end{pmatrix}.
$$
This curve satisfies $\mathcal N_{C_{s_1+s_2}/X_2} \cong \mathcal O 
\oplus \mathcal O(-2)$. Hence, we do not expect that we can retrieve 
its 
information from the PF system.

We have performed similar computations for 3 and 4 parameter cases, all 
of which support this general principle. This leads us to make the

\begin{conjecture}
Let $X$ be a noncompact toric Calabi-Yau threefold with
$\dim H_4(X,\Z)=0$, and say $\{l^1,\dots\,l^m\}$ define $X$ via 
symplectic 
quotient. Let $\omega=\sum_{n>0}c(n,\rho)z^{n+\rho}$ be the generating 
function for
\begin{equation}
\Pi_{ij}^{inst}=\sum_n 
\big(\partial_{\rho_i}\partial_{\rho_j}c(n,\rho)\big)|_{\rho=0}z^n
\end{equation}
with
\begin{equation}
c(n,\rho)^{-1}=\prod_i \Gamma\big(1+\sum_k l^k_i(n_k+\rho_k)\big).
\end{equation}
 If $\mathcal F$ is the prepotential, and 
$s_i=\partial_{\rho_i}\omega|_{\rho=0}$ for each $i$ such that 
$$l^i=
\begin{pmatrix} 
1 & 1 & -1 & -1 & 0 & \dots & 0
\end{pmatrix}
$$
(up to a permutation of the columns of $l^i$), then there are rational 
numbers $m_{ij}\in \mathbb Q$ such that
$$
\sum_{i,j} m_{ij}\Pi_{ij}^{inst}=\sum_i(-1)^{i-1}\frac{\partial 
\mathcal 
F^{inst}}{\partial s_i}.
$$
\end{conjecture}
Here, $\mathcal F^{inst}$ is the instanton part of the prepotential. We 
use the notation $\Pi_{ij}^{inst}$ to distinguish these functions from 
the usual derivatives of $\omega$ (i.e. 
$\Pi_{ij}=\partial_{\rho_i}\partial_{\rho_j}\omega|_{\rho=0}$).

This conjecture is equivalent to the statement that although we cannot 
detect curves with normal bundle $\mathcal O \oplus \mathcal O(-2)$ via 
the $\Pi_{ij}$, we can exhibit all curves with normal bundle  $\mathcal 
O(-1) \oplus \mathcal O(-1)$ using these functions.

We will now exploit this idea, and in so doing discover a way to 
construct a new system of differential operators which provides \it all 
\rm 
mirror symmetry data for this class of examples. 

\subsection{Two building blocks of solutions.}

Assume $X$ is a noncompact Calabi-Yau threefold such that $\dim 
H_4(X,\Z)=0$, and that every two cycle $C \hookrightarrow X$ has normal 
bundle 
$\mathcal O \oplus \mathcal O(-2)$ or $\mathcal O(-1) \oplus \mathcal 
O(-1)$. We will refer to these as $t$ and $s$ curves, respectively, in 
the following. 

Then, as any such space $X$ is obtained by gluing $s$ and $t$ curves 
together in some way, it is reasonable to expect that we can solve all 
these models by extension from the two basic one parameter cases
$$
X_s = \mathcal O(-1) \oplus \mathcal O(-1) \longrightarrow \p^1,
$$
$$
X_t = \mathcal O \oplus \mathcal O(-2) \longrightarrow \p^1.
$$
We have already exhibited the solution on $X_s$. We will now modify 
this slightly to allow extension to the general cases, and subsequently 
demonstrate a similar solution on $X_t$.

Recall, from section 2.1, the differential operator for $X_s$:
$$
\tilde{\mathcal D_1} = \theta_s^2(1-z_s)\theta_s^2.
$$
As before, $\theta_s=z_sd/dz_s$, and $\tilde Y_1=1/(1-z_s)$ is the 
Yukawa coupling. Note that this expression for $\tilde Y_1$ implies a 
classical triple intersection number 1 for $\p^1\hookrightarrow X_s$.

We will now need to make a slightly different choice of Yukawa coupling 
on $X_s$. Recall \cite{W}\cite{GV} that, in the context of the toric 
flop $s \rightarrow -s$, the natural value for the triple intersection 
number is $1/2$. There is a simple proof for this, which we give now. 
From section 2.1, we had the prepotential on local $\p^1$ with 
arbitrary 
triple intersection number (eqn.(\ref{prepotential})):
$$
\mathcal F(z)= K\frac{(\log z)^3}{6}+ \sum_{n >0}\frac{z^n}{n^3}.
$$
A flop of the $\p^1$ on $X_s$ is the same as a change of variables $z 
\longrightarrow 1/z$ in the prepotential:
$$
\mathcal F^{flop}(z)= K\frac{(-\log z)^3}{6}+ \sum_{n 
>0}\frac{z^{-n}}{n^3}.
$$
Taking the difference of these, 
$$
\mathcal F(z)-\mathcal F^{flop}(z)=-\frac{1}{3}K(\log z)^3.
$$
We have ignored the terms including $\sqrt{-1}$, since the Yukawa 
coupling is insensitive to them.
Then according to Witten \cite{W}, we are supposed to find
$$
\mathcal F(z)-\mathcal F^{flop}(z)=-\frac{1}{6}(\log z)^3,
$$
and hence $K=1/2$ is uniquely determined.

This means that we should really be using
$$
Y_1=\frac{1}{2}+\frac{z_s}{1-z_s}
$$
for the Yukawa coupling. We obtain the following differential 
operator 
describing mirror symmetry for $X_s:$
$$
\mathcal D_1= \theta_s^2\Big(\frac{2(1-z_s)}{1+z_s}\Big)\theta_s^2.
$$

Next, let's turn to $X_t$.
Note that the naturally occurring PF operator on the mirror to 
$\mathcal 
O \oplus \mathcal O(-2) \longrightarrow \p^1$, which is
$$
\mathcal D_2'=\theta_t^2-z_t(2\theta_t)(2\theta_t+1),
$$
has no curve information, since there is no double logarithmic 
solution. Moreover, the second Frobenius derivative of the generating 
function 
of solutions has no instanton part. 

Yet, in view of the solution on $X_s$, we can easily exhibit a 
Picard-Fuchs operator for $X_t$; it is given by 
$$
\mathcal D_2 = \partial_t^2(1/Y_2)\partial_t^2.
$$
Here 
$$
t(z)=\log(z_t)+2\sum_{n>0}\frac{(2n-1)!}{{n!}^2}z_t^n
$$
is the mirror map for $X_t$, and $Y_2$ is the Yukawa coupling on $X_t$, 
which in these coordinates is
$$
Y_2=-\frac{1+e^t}{2(1-e^t)}.
$$
The overall negative has no effect on the solution space of the 
differential operator $\mathcal D_2$, but it is taken so that 
$\p^1\hookrightarrow X_t$ has a classical triple self-intersection 
number of $-1/2$. We 
make this choice in analogy with the case $\mathcal O(-3)\rightarrow 
\p^2$, which has intersection number $-1/3$ \cite{KZ}.

Then, as in the compact case, it follows automatically that the 
solutions $\Pi_t$ of $\mathcal D_2\Pi_t=0$ are given as
$$
\Pi_t=\Big(1,t, \frac{\partial \mathcal F}{\partial t}, t\frac{\partial 
\mathcal F}{\partial t}- 2\mathcal F\Big)
$$
where $\mathcal F$ is a holomorphic function in $t$ such that
$$
\frac{\partial^3 \mathcal F}{\partial t^3}=Y_2.
$$

Then it is a simple matter to write down and explicit differential 
operator on the mirror of $X_t$, by a change of coordinates for 
$\mathcal D_2$. We find
\begin{equation}
\label{ournew-2}
\mathcal 
D_2=\theta_t^4-z_t(2\theta_t+2)(2\theta_t+1)^2\theta_t+(z_t)^2(2\theta_t+4)(2\theta_t+3)(2\theta_t+1)2\theta_t,\;\;\;(\theta_{t}:=z_{t}\frac{d}{dz_{t}}).
\end{equation}
The solutions of (\ref{ournew-2}) are generated by the Fr\"obenius 
function:
$$
w(z_{t},\rho):=\sum_{n=0}^{\infty}\frac{1}{\Gamma(1-2n-2\rho)(\Gamma(1+n+\rho))^{2}}(1+\sum_{j=1}^{n}\frac{\rho}{j+\rho})z_{t}^{n+\rho}.
$$
We can easily check that the vector space:
$$ 
\langle \;1,\;t,\;\frac{\d \mathcal F}{\d t},\;2\mathcal F-t\frac{\d 
\mathcal F}{\d t}\;
\rangle_{\C},
$$
is equal to the vector space:
$$ 
\langle 
\;w(z_{t},0),\;\d_{\rho}w(z_{t},0),\;\d^2_{\rho}w(z_{t},0),\;\d^3_{\rho}w(z_{t},0) \;
\rangle_{\C}.
$$
Hence, we have demonstrated the existence of mirror symmetry for both 
$X_s$ and $X_t$, in terms of solutions of new differential operators. 
It 
should be noted that $\mathcal D_1, \mathcal D_2$ cannot be derived 
from any GKZ system on these spaces. 

With these at hand, we can propose a general prescription for local 
mirror symmetry in absence of a 4 cycle. 

\subsection{Mirror Symmetry when $\dim H_4(X,\Z)=0$.}

We can use the results of the previous section to find a general 
solution for such spaces, as follows. From the considerations of 
\cite{I},we 
see that if X a noncompact toric Calabi-Yau threefold with $\dim 
H_4(X,\Z)=0$, then for each $C \in H_2(X,\Z)$, we have
$$
\mathcal N_{C/X}\cong \mathcal O(-1) \oplus \mathcal O(-1) \ \rm or
$$
$$
\mathcal N_{C/X}\cong \mathcal O \oplus \mathcal O(-2).
$$
This is also apparent from the vectors which span the secondary fan. We 
will choose $X$ such that 
$\{C_{s_1},\dots,C_{s_m},C_{t_1},\dots,C_{t_n}\}$ is a basis of 
$H_2(X,\Z)$, where $\mathcal N_{C_{s_i}/X}\cong 
\mathcal O(-1) \oplus \mathcal O(-1), \ \mathcal N_{C_{t_j}/X}\cong 
\mathcal O\oplus \mathcal O(-2)$ $\forall i,j$. Also, let 
$u=(s_1,\dots,s_m,t_1,\dots,t_n).$

From the topological vertex formalism, the authors of \cite{I} were 
able to determine the instanton part of the prepotential for the class 
of 
examples we're considering. Explicitly,
$$
\mathcal F^{inst} 
=\sum_{C_s}\sum_{k>0}\frac{e^{ks}}{k^3}-\sum_{C_t}\sum_{k>0}\frac{e^{kt}}{k^3}. 
$$
Here, the sum over $C_s$ represents the sum over all curves 
$C_s\hookrightarrow X$ such that $N_{{C_s}/X} \cong \mathcal O(-1) 
\oplus \mathcal 
O(-1)$, and similarly for the sum over $C_t$.

As explained in the introduction, our problem reduces to that of 
defining a consistent (triple) 
intersection theory on $X$. Thanks to the simple structure of $X$, 
together with our preliminary choice of intersection numbers for  
$\mathcal 
O(-1) \oplus \mathcal O(-1) \longrightarrow \p^1$ and  $\mathcal O 
\oplus \mathcal O(-2) \longrightarrow \p^1$, there is in fact a unique 
choice. We will first give the general definition, and afterward 
explain its 
significance through an example.

To give the prescription for intersection theory for the general case, 
we will only consider $X$ with toric diagram as in the following 
figure (\ref{chain}).
\begin{figure}[htbp]
\centering
\input{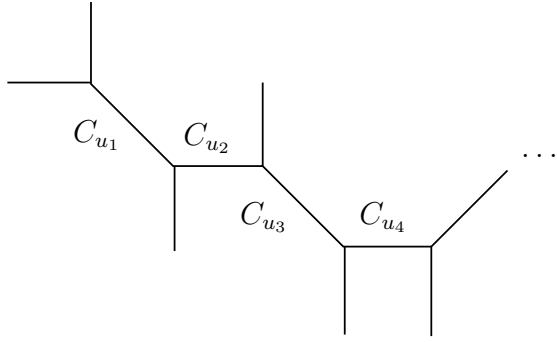}
\caption{$X$ containing a string of curves. }
\label{chain}
\end{figure}
That is, only two curves in $X$ are allowed to intersect at any point. 
Hence, we exclude cases where three curves meet at one point in $X$, 
etc. With this restriction, we can introduce an ordering on the curves 
in 
$X$:
$$
C_{u_1}< C_{u_2} < C_{u_3} < \dots
$$
Define a function
$$
sgn :  H_2(X,\Z)\longrightarrow \{1,-1\}
$$
so that $sgn(C)=1$ if $\mathcal N_{C/X}\cong\mathcal O(-1) \oplus 
\mathcal O(-1)$, and $sgn(C)=-1$ otherwise.

With these conventions, we can now state our conjecture on intersection 
theory.

\begin{definition} Let $X$ be a noncompact toric Calabi-Yau threefold 
such that $\dim H_4(X,\Z)=0$, and suppose $\mathcal N_C \cong \mathcal 
O(-1) \oplus \mathcal O(-1)$ or $\mathcal N_C \cong \mathcal O \oplus 
\mathcal O(-2) \ \forall \ C \in H_2(X,\Z)$. Then the \rm classical 
intersection numbers \it for $X$ are given by
$$
K_{abc}=
$$
\begin{equation}
\label{intersectionnumbers}
\frac{1}{2}\sum_{C \notin \mathcal A}sgn\Big([C_{abc}]+[C]\Big),
\end{equation}
where the sum is taken in homology, and 
$$
[C_{abc}]=[C_{u_a}]+\sum_{C_{u_a}<C_{\alpha}<C_{u_b}}[C_{\alpha}]+[C_{u_b}]+\sum_{C_{u_b}<C_{\beta}<C_{u_c}}[C_{\beta}]+[C_{u_c}].
$$
The sum is taken away from the set 
$$
\mathcal A = \{[C_{u_a}],[C_{u_b}],\dots,[C_{u_a}+C_{u_{a+1}}],\dots\}. 
$$
\end{definition}

\bigskip

This formula can be most simply understood as follows. The curve of 
minimum volume containing all three curves $C_{u_a},C_{u_b}$ and 
$C_{u_c}$
can be represented by the homology class 
$$
[C_{abc}]=[C_{u_a}]+\sum_{C_{u_a}<C_{\alpha}<C_{u_b}}[C_{\alpha}]+[C_{u_b}]+\sum_{C_{u_b}<C_{\beta}<C_{u_c}}[C_{\beta}]+[C_{u_c}].
$$
Then each term of the sum $[C_{abc}]+[C]$ corresponds to a curve in $X$ 
containing $C_{abc}$.

For example, consider the case $a=b=c$. Then both of the sums collapse,  
we are left with only $sgn([C_{u_a}]+[C])$. The sum will contribute 
$\pm 1/2$ for each curve containing $C_{u_a}$, depending on the normal 
bundle of that curve.

Let us now apply this definition to a concrete case. Consider again the 
instanton part of the prepotential from example 1 above, equation 
(\ref{myprepot}):
$$
\mathcal F^{inst}= 
\sum_{n>0}\frac{e^{ns}}{n^3}+\frac{e^{n(s+t)}}{n^3}-\frac{e^{nt}}{n^3}.
$$
Then e.g.
$$
\frac{\partial^3 \mathcal F^{inst}}{\partial 
s^3}=\sum_{n>0}\Big(e^{ns}+e^{n(s+t)}\Big).
$$
Both the $s$ curve and the $s+t$ curve have normal bundle $\mathcal 
O(-1) \oplus \mathcal O(-1)$ (this can be seen from the toric diagram, 
or 
directly from the vectors defining the secondary fan). Thus, each curve 
should have an intersection number equal to $1/2$, which implies
$$
K_{sss}=\frac{1}{2}+\frac{1}{2}=1.
$$
By applying similar reasoning, we obtain the other intersection numbers
$$
K_{tss}=K_{tts}=1/2, \ K_{ttt}=0.
$$
This intersection theory is also compatible with the flop 
$X_1\longrightarrow X_2$ given above. We have also verified that this 
prescription 
gives the simplest possible form for the extended Picard-Fuchs system 
on three parameter models of this type. 

In fact,  using the results of \cite{I}, we can argue that thus 
definition is the right one in general. In \cite{I}, it was shown that 
the 
Gopakumar-Vafa invariants for the spaces of interest are invariant 
under flops; this was done by considering a 3 parameter flop on the 
strip. 
The classical intersection numbers that preserve the polynomial part of 
the prepotential turn to be the ones given above, for the same 3 
parameter models of \cite{I}. Hence, this definition is the unique one 
in 
order to have a theory that transforms sensibly under flops.

\subsection{\bf Extended Picard-Fuchs System for $X_{1}$ and $X_{2}$}

In this subsection, we derive an extended PF system under the 
assumption 
of 
the conjecture given in the previous subsection. First, we look at the 
example $X_{1}$. In this case, we start from four A-model Yukawa 
couplings:  
\begin{eqnarray}
Y_{ttt}&:=&-\frac{e^{t}}{1-e^{t}}+\frac{e^{s+t}}{1-e^{s+t}},\no\\
Y_{tts}&:=&\frac{1}{2}+\frac{e^{s+t}}{1-e^{s+t}},\no\\
Y_{tss}&:=&\frac{1}{2}+\frac{e^{s+t}}{1-e^{s+t}}\no\\
Y_{sss}&:=&1+\frac{e^{s}}{1-e^{s}}+\frac{e^{s+t}}{1-e^{s+t}}.
\end{eqnarray}
The constant part of each Yukawa coupling is given by the conjecture, 
and the instanton (nonconstant) part was taken from \cite{I}. We repeat 
here the PF operators given by the standard toric construction of the 
mirror 
manifold $\hat{X_{1}}$:  
\begin{eqnarray}
\label{PFx1}
\mathcal 
D_1&=&\theta_1(\theta_1-\theta_2)-z_1(2\theta_1-\theta_2)(2\theta_1-\theta_2+1),\no\\
\mathcal 
D_2&=&(2\theta_1-\theta_2)\theta_2-z_2(\theta_1-\theta_2)\theta_2,
\no\\
\mathcal D_3&=&\theta_1\theta_2-z_1z_2(2\theta_1-\theta_2)\theta_2.
\end{eqnarray}
By solving (\ref{PFx1}), we obtain mirror maps $s=s(z_{1},z_{2})$ and 
$t=t(z_{1},z_{2})$. In particular, the Jacobian of these mirror maps 
are 
written in terms of
simple functions, as follows:
\begin{eqnarray}
\frac{\d t}{\d u_1} = 
\frac{1}{\sqrt{1-4z_1}},\;\;\;\;
\frac{\d t}{\d u_2} := 0,
\;\;\;\;
\frac{\d s}{\d u_1} = \frac{1}{2}\frac{-1+4z_1+\sqrt{1-4z_1}}{-1+4z_1},
\;\;\;\;
\frac{\d s}{\d u_2} = 1,
\label{jx1}
\end{eqnarray}
where $u_{i}=\log(z_{i})$. With this data, we can compute the $B$ model 
Yukawa couplings in $u^{i}$ coordinates, and they turn out to be 
rational functions in $z_{i}$ whose denominators are given by the 
divisor 
of the defining equation of discriminant locus of $\hat{X_{1}}$:
\begin{equation}
dis(\hat{X_{1}})=(1-z_2+z_1z_2^2)(1-4z_{1}).
\end{equation}
The explicit results are given as follows:  
\begin{eqnarray}
Y_{111} &=& 
-\frac{1}{2}\frac{z_1(-4z_1+5-7z_2+12z_1z_2+2z_2^2-5z_1z_2^2+
4z_1^2z_2^2)}{(1-z_2+z_1z_2^2)(-1+4z_1)^2},\no\\
Y_{112} &=& 
-\frac{1}{2}\frac{(1-2z_1-z_2+4z_1z_2-z_1z_2^2+2z_1^2z_2^2)}
{(1-z_2+z_1z_2^2)(-1+4z_1)},\no\\
Y_{122} &=& 
-\frac{1}{2}\frac{(-1+z_2+z_1z_2^2)}{(1-z_2+z_1z_2^2)},\no\\
Y_{222} &=& \frac{1-z_1z_2^2}{1-z_2+z_1z_2^2}.\no\\
\label{yux1}
\end{eqnarray}
These  results show that the conjecture given in the previous section 
is compatible with Conjecture 1 in Section 1. Therefore, we can 
construct an
extended PF system by using the strategy outlined in Section 2. For 
brevity, 
we introduce here the following notation:
\begin{equation}
M_{a}^{\alpha}(t_{*}):=\sum_{b}(Y_{a}^{-1}(t_{*}))^{\alpha 
b}\d_{a}\d_{b}
\psi_{0}.
\label{mmat}
\end{equation}
In the case of $X_{1}$,  
we have two integrability conditions given in (\ref{1}):
\begin{equation}
M_{1}^{1}(t,s)=M_{2}^{1}(t,s),\;\;M_{1}^{2}(t,s)=M_{2}^{2}(t,s),
\end{equation} 
where we use the subscript $1$ and $2$ for $t$ and $s$.
By explicit computation, these two conditions turn out to be the
same, and translated into a differential equation in $z_{i}$ variables 
by using (\ref{jx1}) and (\ref{yux1}), we obtain: 
\begin{equation}
\bigl((1+z_2+z_1z_2^{2})\mathcal D_{1}+\mathcal D_{2}+z_2\mathcal D_{3}
\bigr)\psi_{0}=0.
\label{1x1}
\end{equation}
Next, we consider the second integrability condition given in 
(\ref{2}):
\begin{equation}
\d_{1}M_{1}^{1}(t,s)=\d_{2}M_{2}^{2}(t,s),\;\;\d_{2}M_{1}^{1}(t,s)=0,\;\;
\d_{1}M_{2}^{2}(t,s)=0.
\end{equation}
By explicit computation, we found that the second and the third
conditions are translated into 
one rational differential equation:
\begin{equation}
\theta_{2}\mathcal D_{1}\psi_{0}=0.
\label{2x1}
\end{equation}
The first condition is also translated into a rational differential 
equation
but the result is very complicated.
Now, we assert that (\ref{1x1}) and (\ref{2x1}) are a minimal set of 
extended 
PF operators for $X_{1}$. The reason is the following. Let us consider 
the 
large radius limit of (\ref{1x1}) and (\ref{2x1});
\begin{equation}
(\theta_{1}^2+\theta_1\theta_2-\theta_2^2)\psi_{0}=0,\;\;
(\theta_1^2\theta_2-\theta_1\theta_2^2)\psi_{0}=0.
\end{equation} 
These conditions are equivalent to the relations of the classical 
cohomology 
ring of $\overline{X}_{1}$:
\begin{equation}
k_{t}^2+k_tk_{s}-k_s^2=0,\;\;k_{t}^2k_s-k_tk_s^2=0,
\end{equation}
which reproduces the conjectured triple intersection numbers, up to an
overall 
scaling. From this fact, we can see that  (\ref{1x1}) and (\ref{2x1}) 
give 
us a complete set of relations for the 
classical cohomology ring of $\overline{X}_{1}$ at the large 
radius limit. Since the PF equations are nothing but the 
non-commutative 
version
of the relations of the quantum cohomology ring of $\overline{X}_{1}$, 
which 
reduce to relations of classical cohomology at the large radius limit, 
\cite{guest}, we can propose the following set of differential 
operators
as an extended PF system: 
\begin{eqnarray}
\tilde{\mathcal D}_{1}&=&(1+z_2+z_1z_2^{2})\mathcal D_{1}+\mathcal 
D_{2}+z_2
\mathcal D_{3},\no\\
\tilde{\mathcal D}_{2}&=&\theta_{2}\mathcal D_{1}.
\label{ex1}
\end{eqnarray}
We checked that the solution space of (\ref{ex1}) is given by,
\begin{equation}
\langle\; 1,\;t,\;s,\;\frac{\d \mathcal F}{\d t},\;
\frac{\d \mathcal F}{\d s},\;
2\mathcal F-t\frac{\d \mathcal F}{\d t}-s\frac{\d \mathcal F}{\d s}
\;\rangle_{\C}.
\end{equation}
Of course, we can derive the B-model Yukawa couplings (\ref{yux1}) by 
using 
(\ref{ex1}) as the starting point. An explicit example of this kind of 
computation will be given in Section 6 of this paper.  

We can also construct an extended PF system of $X_{2}$ in the same way 
as 
$X_{1}$. Here, we briefly present the data of this construction. 
The starting 
point is the A-model Yukawa couplings:   
\begin{eqnarray}
Y_{111}&:=&\frac{e^{s_1}}{1-e^{s_1}}-\frac{e^{s_1+s_2}}{1-e^{s_1+s_2}},\no\\
Y_{112}&:=&-\frac{1}{2}-\frac{e^{s_1+s_2}}{1-e^{s_1+s_2}}\no\\
Y_{122}&:=&-\frac{1}{2}-\frac{e^{s_1+s_2}}{1-e^{s_1+s_2}},\no\\
Y_{222}&:=&\frac{e^{s_2}}{1-e^{s_2}}-\frac{e^{s_1+s_2}}{1-e^{s_1+s_2}},
\end{eqnarray}
and the ordinary PF operators:
\begin{eqnarray}
\mathcal 
D_{1}&:=&\theta_{1}^{2}-\theta_{1}\theta_{2}-z_1(\theta_{1}+\theta_{2})(\theta_{1}-\theta_{2}),\no\\
\mathcal 
D_{2}&:=&\theta_{2}^{2}-\theta_{1}\theta_{2}-z_2(\theta_{2}+\theta_{1})(\theta_{2}-\theta_{1}),\no\\
\mathcal 
D_{3}&:=&\theta_{1}\theta_{2}-z_1z_2(\theta_{1}+\theta_{2}+1)(\theta_{1}+\theta_{2}).
\label{pfx2}
\end{eqnarray}
Let us introduce the logarithm of the $B$ model coordinates $z_{i}$.
\begin{equation}
u_1=\log(z_1),\;\;u_2=\log(z_2).
\end{equation}
By solving (\ref{pfx2}), we obtain the mirror maps 
$s_{1}=s_{1}(z_{1},z_{2})$ and $s_{2}=s_{2}(z_{1},z_{2})$ and their 
Jacobian: 
\begin{eqnarray}
\frac{\d s_1}{\d u_1} 
&=& \frac{1}{2}\frac{(-\sqrt{1-4z_1z_2}-1+4z_1z_2)}{(4z_1z_2-1)},\;\;
\frac{\d s_1}{\d u_2} 
= -\frac{1}{2}\frac{(-1+4z_1z_2+\sqrt{1-4z_1z_2})}{(4z_1z_2-1)},\no\\
\frac{\d s_2}{\d u_1} 
&=& -\frac{1}{2}\frac{(-1+4z_1z_2+\sqrt{1-4z_1z_2})}{(4z_1z_2-1)},\;\;
\frac{\d s_2}{\d u_2} 
= \frac{1}{2}\frac{(-\sqrt{1-4z_1z_2}-1+4z_1z_2)}{(4z_1z_2-1)}.
\end{eqnarray}
With this data, we can compute the $B$ model Yukawa couplings in 
$u^{i}$: 
\begin{eqnarray}
Y_{111} &=& \frac{1}{2}z_1\frac{(5z_1z_2-2-12z_1z_2^2+7z_2+4z_2^3z_1-
5z_2^2-4z_1^2z_2^2)}{(z_2-1+z_1)(4z_1z_2-1)^2},\no\\
Y_{112} &=& 
\frac{1}{2}\frac{(1-z_1-z_2-z_1z_2-z_2z_1^2+z_1z_2^2+4z_1^2z_2^2+
4z_2^2z_1^3-4z_2^3z_1^2)}{(z_2-1+z_1)(4z_1z_2-1)^2},\no\\
Y_{122} &=& 
\frac{1}{2}\frac{(1-z_2-z_1-z_2z_1-z_1z_2^2+z_2z_1^2+4z_2^2z_1^2+
4z_1^2z_2^3-4z_1^3z_2^2)}{(z_1-1+z_2)(4z_2z_1-1)^2},\no\\
Y_{222} 
&=&\frac{1}{2}z_2\frac{(5z_2z_1-2-12z_2z_1^2+7z_1+4z_1^3z_2-5z_1^2-
4z_2^2z_1^2)}{(z_1-1+z_2)(4z_2z_1-1)^2},
\end{eqnarray}
and they turn out to be rational functions in $z_{i}$ whose  
denominators 
are divisors of defining equation of discriminant locus of 
$\hat{X}_{2}$:
\begin{equation}
dis(\hat{X}_{2})=(1-z_{1}-z_{2})(1-4z_1z_2).
\end{equation}
The derivation of the extended PF system by using the recipe in Section 
2 proceeds 
in the same way as $X_{1}$. In this case, we only have to consider
\begin{equation}
M_{1}^{1}(s_1,s_2)=M_{2}^{1}(s_1,s_2),\;\;M_{1}^{2}(s_1,s_2)=M_{2}^{2}
(s_1,s_2),
\label{1x2}
\end{equation} 
and 
\begin{equation}
\d_{1}M_{1}^{1}(s_1,s_2)=\d_{2}M_{2}^{2}(s_1,s_2),\;\;\d_{2}M_{1}^{1}
(s_1,s_2)=0,\;\;
\d_{1}M_{2}^{2}(s_1,s_2)=0.
\label{2x2}
\end{equation} 
(\ref{1x2}) gives us one differential equation for $\psi_{0}$ 
with rational function coefficients in $z_{i}$:
\begin{equation}
\bigl(\mathcal D_{1}+\mathcal D_{2}+
(1+z_1+z_2)\mathcal D_{3}\bigr)\psi_{0}=0.
\end{equation}
 As for (\ref{2x2}),
the second and the third conditions give us a differential equations 
for $\psi_{0}$:
\begin{equation}
(\theta_{1}-\theta_{2})\mathcal D_{3}\psi_{0}=0,
\end{equation} 
and the first one gives us a complicated rational differential 
equation.
For the same reasoning as $X_{1}$, 
we can propose an extended PF system for $X_{2}$ as 
follows: \begin{eqnarray}
&&\tilde{\mathcal D}_{1}:=\mathcal D_{1}+\mathcal D_{2}+
(1+z_1+z_2)\mathcal D_{3},\no\\
&&\tilde{\mathcal D}_{2}:=(\theta_{1}-\theta_{2})\mathcal D_{3}.
\end{eqnarray}
We have also constructed an extended PF system for a three parameter 
space $X_{3}$, in order to further test the conjecture made in the 
previous 
section. Specifically, $X_3$ satisfies $\dim H_2(X,\Z)=3$, $\dim 
H_4(X,\Z)=0$, and is defined by the following vectors:
\begin{equation}
\begin{pmatrix}
 l^1 \\ l^2 \\ l^3  
\end{pmatrix}=
\begin{pmatrix}
	1 & 1 & -1 & -1 & 0 & 0 \\ 0 & -1 & -1 & 1 & 1 & 0 \\ 0 & -1 & 1 & 0 & 
-1 & 1
 \end{pmatrix}.
\end{equation} 
The results are collected in Appendix A.

\section{Adding open strings to $\mathcal O(-1)\oplus \mathcal 
O(-1)\rightarrow \p^1$.}

So far, we have been able to demonstrate the existence of new 
differential operators which determine mirror symmetry for noncompact 
toric 
Calabi-Yau manifolds which have no 4 cycle. One may also wonder what 
the 
applications are to the PF system derived for open mirror symmetry 
\cite{LM}\cite{F}. We will see that again, some modification of open 
string 
PF operators is necessary.

\subsection{Review of open string geometry.}

Recall \cite{AV}\cite{F} that open string mirror symmetry is a local 
isomorphism of moduli spaces $(\mathcal X, \mathcal L)$ and $(\mathcal 
{\hat X}, 
\mathcal C)$; $X$ and $\hat X$ are Calabi-Yau manifolds which are 
mirror in 
the usual sense, and $L\subset X$ is Lagrangian, while $C\subset \hat 
X$ is 
holomorphic. In the case at hand, $(\mathcal X, \mathcal L)$ will be 
given by
$$
X_r=\{(w_1,\dots,w_4)\in \C^4-Z : 
|w_1|^2+|w_2|^2-|w_3|^2-|w_4|^2=r\}/S^1,
$$
together with either
\begin{equation}
\label{lag1}
L_{r,c}=X_r\cap \{|w_2|^2-|w_4|^2=c, \ |w_3|^2-|w_4|^2=0, \ \sum_i 
\arg(w_i)=0\}
\end{equation}
or
\begin{equation}
\label{lag2}
L_{r,c}'=X_r\cap \{|w_2|^2-|w_4|^2=0, \ |w_3|^2-|w_4|^2=c, \ \sum_i 
\arg(w_i)=0\}.
\end{equation}
There are then local moduli space isomorphisms $(\mathcal X, \mathcal 
L)\cong(\mathcal {\hat X}, \mathcal C)$, $(\mathcal X, \mathcal 
L')\cong(\mathcal {\hat X}, \mathcal C')$, where
$$
\hat X_{z_1}=\{(u,v,y_1,y_2)\in \C^2\times (\C^*)^2: 
uv+1+y_1+y_2+z_1y_1y_2^{-1}=0\},
$$
\begin{equation}
\label{hol1}
C_{z_1,z_2}=\hat X_{z_1}\cap \{y_2^{-1}y_1=z_2, \ y_1=1\},
\end{equation}
\begin{equation}
\label{hol1}
C_{z_1,z_2}'=\hat X_{z_1}\cap \{y_2^{-1}y_1=1, \ y_1=z_2\}.
\end{equation}
The detailed derivation of these spaces is given in \cite{AV}. One 
mathematical implication of open string mirror symmetry is that the 
geometry of $(\hat X,C)$ should determine the genus 0 open 
Gromov-Witten invariants 
of $(X,L)$. This means that there should be functions defined on the 
moduli space  $(\mathcal {\hat X}, \mathcal C)$ which count holomorphic 
maps 
$f:D=\{z\in \C:|z|\le 1\}\rightarrow X$ such that $f(\partial D)\subset 
L$. Furthermore, in many cases such functions can be derived from an 
open 
string Picard-Fuchs system on $(\hat X,C)$. However, for the case at 
hand, 
it will be shown that the same sort of modification of the PF system, 
proposed for ordinary closed mirror symmetry, is also necessary in the 
open string setting. 

 We turn to the PF system on the moduli spaces $(\mathcal {\hat X}, 
\mathcal 
C)$ and $(\mathcal {\hat X}, \mathcal C')$.

\subsection{Moduli space and Picard-Fuchs system for $(\mathcal {\hat 
X}, 
\mathcal C)$.}

From \cite{F}, we can define ``open period integrals" on $(\mathcal 
{\hat X}, 
\mathcal C)$ by
\begin{equation}
\label{openperiods}
\Pi_{\Gamma}(z_1,z_2)=\int_{\Gamma}\frac{dudvdy_1dy_2/(y_1y_2)}{(uv+1+y_1+y_2+z_1y_1y_2^{-1})(y_1-z_2y_2)(y_1-1)},
\end{equation}
$\Gamma \in H_4(\C^2\times (\C^*)^2-\hat X,\hat X-C,\Z)$.
Then the derivations in \cite{F} lead to a Picard-Fuchs system for 
$(\mathcal {\hat X}, \mathcal C)$, given as
\begin{equation}
\label{openPF}
{\mathcal 
D}_1=\theta_1(\theta_1+\theta_2)-z_1\theta_1(\theta_1+\theta_2),
\end{equation}
$$
{\mathcal 
D}_2=\theta_2(\theta_1+\theta_2)-z_2\theta_2(\theta_1+\theta_2), 
$$
$$
\theta_i=z_i\frac{d}{dz_i};
$$
these operators satisfy ${\mathcal D}_i\Pi_{\Gamma}(z_1,z_2)=0$.
This is exactly the noncompact PF system of the vectors 
$l^1=(1,1,-1,-1,0,0), \ l^2=(0,1,0,-1,1,-1)$, and agrees with the 
results of 
\cite{LM}.

As was shown in \cite{CKYZ}, the solution space of $\{{\mathcal 
D}_1,{\mathcal D}_2\}$ can be obtained, using the Frobenius method, 
from a 
function
$$
\omega_0(z,\rho)=\sum_{n \ge 
0}c(n,\rho)z_1^{n_1+\rho_1}z_2^{n_2+\rho_2},
$$
where
$$
c(n,\rho)^{-1}=\Gamma(1+n_1+\rho_1)\Gamma(1+n_1+\rho_1+n_2+\rho_2)\Gamma(1-n_1-\rho_1)*
$$
$$
\Gamma(1-n_1-\rho_1-n_2-\rho_2)\Gamma(1+n_2+\rho_2)\Gamma(1-n_2-\rho_2).
$$
According to \cite{LMW}, the solutions are expected to be 
$$
(1,t_1,t_2,W_1,W_2,\dots).
$$ 
$t_1$ and $t_2$ give the open string mirror map, and this is trivial 
for the present example, so we have $t_i(z)=log(z_i)$. Also, the $W_i$ 
count discs on $(X,L)$.

Upon looking at the equations of $(X,L)$, we can make the following 
geometric observation about a map $f:D\rightarrow X$ with $f(\partial 
D)\subset L$. In the region where $w_3=w_4=0$, $L$ will intersect the 
$\p^1$ of $X$; hence, such an $f$ must obey $f(D)\subset \p^1$. Then 
the 
natural interpretation of the variable $z_2$ is as a parameter 
controlling 
the size of a holomorphic disc $D\hookrightarrow X$. It is therefore 
expected that one of the double logarithmic solutions of (\ref{openPF}) 
will look like
\begin{equation}
\label{opensuper1}
W_1(z_2)=\sum_{n>0}\frac{z_2^n}{n^2},
\end{equation}
where the log terms have been disregarded due to ambiguity \cite{AV}. 
And indeed, it is the case that
$$
W_1(z_2)= (\partial_{\rho_2}^2\omega_0)|_{\rho=0}.
$$
The problem, though, is that $(\partial_{\rho_2}^2\omega_0)|_{\rho=0}$ 
is not a solution of (\ref{openPF}). The easiest way to see this is to 
note that $W_1$ is independent of $z_1$, and (\ref{openPF}) reduces to 
${\mathcal D}_2=(1-z_2)\theta_2^2$ if $z_1=0$.

The minimal resolution of this issue, which continues in the spirit of 
raising the power of PF operators, is to instead work with the system
\begin{equation}
\label{openextenz}
\{{\mathcal D}_1,\theta_2{\mathcal D}_2\}.
\end{equation}
$W_1$ is indeed a solution of these higher order operators. 

A lingering difficulty, even after such a PF extension, is that (as 
will be shown below) there is expected to be another disc counting 
function
$$
W_2(z_1,z_2)=\sum_{n>0}\frac{(z_1/z_2)^n}{n^2},
$$ 
which measures the size of the ``other disc"; that is, since $L$ splits 
$\p^1$ into two discs, there should be a function corresponding to each 
disc. Functions with essential singularities such as $W_2$ are not 
allowed to be Picard-Fuchs solutions.

 One way around this is to instead use the vectors $l^1,-l^2$ to write 
down the noncompact PF system. Then it is easy to see that the 
resulting extended system
$$
\{\tilde {\mathcal D}_1,\theta_2 \tilde {\mathcal D}_2\}
$$
will have solutions
\begin{equation}
\label{newper}	
	\tilde W_1(z_2)=\sum_{n>0}\frac{z_2^n}{n^2}, \ \ \ \ \ \tilde W_2(z_1, 
z_2)= \sum_{n>0}\frac{(z_1z_2)^n}{n^2}.
\end{equation}
These are the same as found in \cite{OV}. 

Similarly, we can perform calculations on the family $(\hat X,C')$. 
This 
moduli space is given by vectors $k^1=l^1, \ k^2=(0,0,1,-1,1,-1)$, and 
the 
open string PF system we arrive at is
$$
{\mathcal D}'_1=\theta_1^2-z_1(\theta_1-\theta_2)(\theta_1+\theta_2),
$$
$$
{\mathcal 
D}'_2=(\theta_2-\theta_1)\theta_2-z_2(\theta_1+\theta_2)\theta_2.
$$
Again, analogously to the above, let $\omega'_0$ be the generator of 
solutions of $\{ {\mathcal D}'_1, {\mathcal D}'_2\}$. Then there is a 
disc counting function
\begin{equation}
\label{bigdisc}
\partial_{\rho_2}^2\omega'_0|_{\rho=0}=W'(z_1,z_2)=\sum_{n_2>n_1\ge0}\frac{(-1)^{n_1}(n_1+n_2-1)!}{(n_1!)^2(n_2-n_1)!n_2}z_1^{n_1}z_2^{n_2}	
\end{equation}
which agrees with the result of \cite{AV}. Yet, once again we have the 
problem of this not being a solution of the given PF system; the same 
modification gives the system
$$
\{{\mathcal D}'_1,\theta_2{\mathcal D}'_2\}.
$$
$W'$ is indeed among the solutions of this.

The moral of this discussion is that the open string PF system, 
constructed in \cite{LM}, is also incomplete in certain cases. Though 
the 
geometric meaning of raising the power of operators is less clear this 
time, we find that the same techniques are effective in open and closed 
string calculations. Moreover, the extension $\mathcal D_2 \rightarrow 
\theta_2 \mathcal D_2$ (rather than $\mathcal D_1 \rightarrow \theta_1 
\mathcal D_1$) is the natural one. This follows because $z_2$ is the 
open 
string variable, and $(\partial_{\rho_2}^2\omega_0)|_{\rho=0}$ counts 
discs; hence, we must assure that the second partial derivative in 
$\rho_2$ is a solution of the system. 

 Next, we will give an open string period integral definition for these 
degenerate situations.

\subsection{Period integrals for $(\mathcal {\hat X}, \mathcal C)$.}

So far, it has been seen that the open PF system found in \cite{LM}, 
which was later shown to be derived from a set of period integrals 
\cite{F}, does not always give the disc-counting functions one is 
interested 
in. Since local mirror symmetry on \loc yields an incomplete PF system, 
it is not so surprising that open strings on this same space should 
exhibit a similar failing. 

Hence, we need a definition of open string period integrals. For 
motivation, let's review some geometric facts about $(\mathcal {\hat 
X}, \mathcal 
C)$. Let $y$ be a local coordinate on $\Sigma$. Then following 
\cite{AV}, 
we can think of the curve $C$ as
$$
C_{z_1,z_2}=\hat X_{z_1}\cap\{v=0, y=z_2\}= \C\times \{z_2\in \Sigma\}.
$$ 
Then the coordinate on  $C\cong \C$ is $u$, and $z_2$ parameterizes a 
family of curves in $\Sigma$. 

\begin{figure}[htbp]
\centering
\input{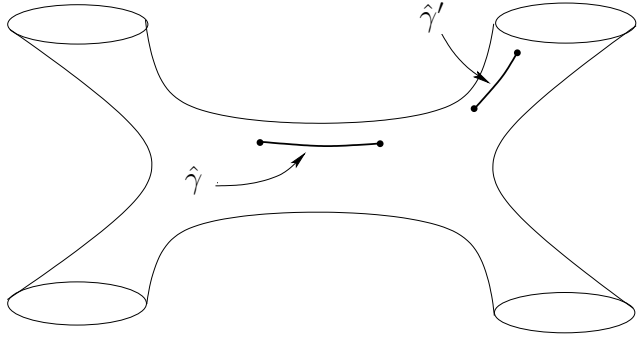}
\caption{Real curves defining open string periods on $\Sigma$. }
\label{thisone}
\end{figure}

Earlier, it was noted that the problem of period integrals was 
reducible to that of integrals on $\Sigma$. Here it is beneficial to 
make the 
same simplification. Notice that, when projected to $\Sigma$, the 
family 
of curves $\{C_{z_1,y}|y\in[z,z_2]\}$ becomes a real curve connecting 
$z$ to $z_2$. Hence, the sensible extension of Definition 1 to open 
strings is  
\begin{definition}
\label{openperiodsforyou} Let $\hat X, \Sigma$ be as given in 
Definition 1, 
and $C$ as above. Choose $z, z_2 \in \Sigma$ and $\hat{\gamma} \in 
H_1(\Sigma,\{z,z_2\},\Z)$. Then the \rm open period integrals  \it of 
$(\mathcal {\hat X}, \mathcal C)$ are defined to be
$$
W(z_1,z_2)=\int_{\hat \gamma} 
\Res_{f=0}\Big(\frac{dy_1dy_2/(y_1y_2)}{f(z_1,y_1,y_2)}\Big).
$$
\end{definition}
For the purposes of the definition, $z$ is considered to be fixed on 
$\Sigma$, and $z_2$ is taken as a parameter.
In the local $\p^1$ example, the relevant curves $\hat \gamma$ are 
shown in Figure \ref{thisone}.
We now move on to the evaluation of these, and show agreement with the 
solutions of the proposed extended open string PF system of the last 
section. 

There are two integrals, associated to the curves of the figure, and 
their calculation proceeds as follows.
$$
W(z_1,z_2)=\int_{\hat{\gamma}}\log y_1\frac{dy_2}{y_2}=\int_{z}^{z_2} 
\log \Big( \frac{-1-y_2}{1+z_1y_2^{-1}}\Big)\frac{dy_2}{y_2}=
$$
$$
\sum_{n>0}\frac{(-z_2)^n}{n^2}-\sum_{n>0}\frac{(-z_2^{-1}z_1)^n}{n^2},
$$
which is what we saw in the previous section from the PF system (after 
the rotation $y_2\rightarrow e^{i\pi}y_2$). If one is so inclined, the 
functions of (\ref{newper}) can be reproduced in the same way; this 
simply amounts to a change of the coordinate choices made in the mirror 
construction.

We also find
$$
W'(z_1,z_2)=\int_{{\hat{\gamma}}'}\log 
y_2\frac{dy_1}{y_1}=\int_{z}^{z_2}\log\Big(\frac{-1-y_1 - 
\sqrt{(1+y_1)^2-4z_1y_1}}{2} 
\Big)\frac{dy_1}{y_1}.
$$
This integral is more difficult to directly evaluate, but as in 
\cite{AV} we can simply note that
$$
z_2\frac{d}{dz_2}W'=\log\Big(\frac{-1-z_2 - 
\sqrt{(1+z_2)^2-4z_1z_2}}{2} \Big).
$$
After Taylor expanding about $z_1=0$ and integrating the result in 
$z_2$, this matches (\ref{bigdisc}).

\section{More general local geometries.}

We have come a long way toward a more complete picture of the 
differential equations governing local mirror symmetry. However, we 
have yet to 
test these ideas in the domain of applicability of \cite{CKYZ}; namely, 
local Calabi-Yau manifolds $K_S$, where $S$ is a Fano surface and $K_S$ 
is its canonical bundle. The system of differential operators as given 
in \cite{CKYZ} cannot be complete, since it is not possible to fix the 
prepotential simply by analyzing the solution set of the system. We 
will find that our techniques still apply, though a unified treatment 
for 
all examples is less clear.

\subsection{One 1-parameter example.}

The simplest, though rather trivial, example is  $\can = \mathcal O(-3) 
\rightarrow \p^2$. This can be defined as a symplectic quotient
$$
\can = \{(w_1,\dots,w_4)\in \C^4-Z : 
|w_1|^2+|w_2|^2+|w_3|^2-3|w_4|^2=r\}/S^1,
$$
where $Z=\{w_1=w_2=w_3=0\}$, $r \in \R^+$ and the $S^1$ action is given 
as 
$$(w_1,\dots,w_4)\rightarrow 
(e^{i\theta}w_1,e^{i\theta}w_2,e^{i\theta}w_3,e^{-3i\theta}w_4).$$

The original paper \cite{CKYZ} associates a Picard-Fuchs system here, 
which is ultimately derivable from the period integrals of the mirror
$$
\hat X_z=\{(u,v,y_1,y_2)\in \C^2\times (\C^*)^2: 
uv+1+y_1+y_2+zy_1^{-1}y_2^{-1}=0\}.
$$
Again, from \cite{H}:
$$
\Pi_{\Gamma}(z)=\int_{\Gamma}\frac{dudvdy_1dy_2/(y_1y_2)}{uv+1+y_1+y_2+zy_1^{-1}y_2^{-1}}
$$
for $\Gamma \in H_4(\C^2\times (\C^*)^2-\hat X,\Z)$ are the period 
integrals. Then we immediately recover the well known PF operator
$$
\mathcal D= \theta^3+z(3\theta)(3\theta+1)(3\theta+2), \ \ \ 
\theta=z\frac{d}{dz},
$$
whose solution space is generated by a function 
$\omega_0(z,\rho)=\sum_{n\ge0}c(n,\rho)z^{n+\rho}$. Here, the 
coefficients can be written
$$
c(n,\rho)=(\Gamma(1-3n-3\rho)\Gamma(1+n+\rho)^3)^{-1}.
$$
If we write the solutions in the variable 
$t=\partial_{\rho}\omega_0|_{\rho=0}$, we get
$$
\Pi=(1,t,\partial \mathcal F/ \partial t)
$$
Naturally, this implies that in the $t$ variable, it must be the case 
that 
$$
\mathcal D= \partial_t\Big(\frac{\partial^3 \mathcal F}{\partial 
t^3}\Big)^{-1}\partial_t^2.
$$
Then, we can again give a `compactified' operator
$$
\partial _t \mathcal D
$$
which possesses a completed set of solutions. It is actually equivalent 
to just work with
$$
\theta \mathcal D=\theta(\theta^3+z(3\theta)(3\theta+1)(3\theta+2))
$$
on account of the invertibility of the Jacobian. And this is our new, 
extended PF operator.

\subsection{Completing mirror symmetry for Hirzebruch surfaces.}

One parameter spaces of type $K_S$ have already been exhausted, by the 
$K_{\p^2}$ case. We will now turn to the two parameter spaces, namely 
the canonical bundle over the Hirzebruch surfaces $F_0,F_1,F_2$. As is 
well-known \cite{CKYZ}, the instanton part of the double log solution 
of the 
standard 
PF system is given by a linear combination of the
$\frac{\d\mathcal F_{inst.}}{\d t_{a}}$. This fact tells us that the 
standard 
PF system of $K_{S}$ already includes the information coming from 
$Y_{a}^{-1}(t_{*})$ in 
(\ref{0}). Therefore, we can take a short cut in the process of 
constructing an
extended 
PF system on $K_{S}$. Examples of this explicit 
construction 
will be given in the next subsection.

The symplectic quotient description is given by $K_{F_n}=$
$$
\{-2|w_1|^2+|w_2|^2+|w_3|^2=r_1^n, \ 
(-2+n)|w_1|^2-n|w_2|+|w_4|^2+|w_5|^2=r_2^n\}/(S^1)^2
$$
where $(w_1,\dots,w_5)\in \C^5-Z_n$. That is, the vectors in the 
secondary fan are

$$
\begin{pmatrix}
 l^1_n \\ l^2_n   
\end{pmatrix}=
\begin{pmatrix}
	-2 & 1 & 1 & 0 & 0  \\ -2+n & -n & 0 & 1 & 1 
 \end{pmatrix}.
$$
The methods of \cite{CKYZ} lead to PF operators:\\
$K_{F_0}:$
$$
\mathcal 
D_1^0=\theta_1^2-z_1(2\theta_1+2\theta_2)(2\theta_1+2\theta_2+1),
$$
$$
\mathcal 
D_2^0=\theta_2^2-z_2(2\theta_1+2\theta_2)(2\theta_1+2\theta_2+1)
$$
$
K_{F_1}:
$
$$
\mathcal 
D_1^1=\theta_1(\theta_1-\theta_2)-z_1(2\theta_1+\theta_2)(2\theta_1+\theta_2+1),
$$
$$
\mathcal D_2^1=\theta_2^2-z_2(2\theta_1+\theta_2)(\theta_1-\theta_2)
$$
$
K_{F_2}:
$
$$
\mathcal D_1^2=\theta_1(\theta_1-2\theta_2)-z_12\theta_1(2\theta_1+1),
$$
$$
\mathcal 
D_2^2=\theta_2^2-z_2(2\theta_2-\theta_1)(2\theta_2-\theta_1+1).
$$

For each respective system, we let  $t_1^n,t_2^n$ be the logarithmic 
solutions. Each case comes equipped with a single double log solution 
$W_n$. If $\omega^n$ is the generating function of solutions on 
$K_{F_n}$ 
and $\Pi^n_{ij}=\partial_{\rho_i}\partial_{\rho_j}\omega^n|_{\rho=0},$ 
then we can write these as
$$
W_0=\Pi^0_{12}, \ W_1=\Pi^1_{11}+2\Pi^1_{12}, \  
W_2=\Pi^2_{11}+\Pi^2_{12}.
$$
Taking $\mathcal F_n$ for the prepotential on $K_{F_n}$, we have the 
following equalities:
$$
W_n=2\frac{\partial \mathcal F_n}{\partial t_1^n}+(2-n)\frac{\partial 
\mathcal F_n}{\partial t_2^n}.
$$

By a comparison of power series, we can demonstrate that the 
$\Pi^n_{ij}$ contain all the information necessary to derive the 
(instanton part 
of) the prepotential on $K_{F_n}$. We find
\begin{equation}
\begin{pmatrix}
\Pi_{11}^n \\ \Pi_{12}^n 
\end{pmatrix}=
\begin{pmatrix}
	0 & 4 \\ 2 &
 -3n+2 
\end{pmatrix}
\begin{pmatrix}
\partial \mathcal F_n/\partial t_1^n \\ \partial \mathcal F_n^/\partial 
t_2^n
\end{pmatrix}.
\end{equation}
The equality above holds at the level of instanton parts of $\mathcal 
F_n$. We will now investigate the classical terms of these 
prepotentials.

\subsection{B-model Yukawa Coupling of $K_{F_{n}}$}

First, we give here the discriminant locus of ${K}_{F_{n}}$ where the
corresponding mirror hypersurface $\hat{K}_{F_{n}}$ becomes singular: 
\begin{eqnarray}
dis(\hat{\bf K}_{{\bf F}_{0}})&=&1-8(z_1+z_2)+16(z_1-z_2)^{2},\\
dis(\hat{\bf K}_{{\bf F}_{1}})&=&(1-4z_1)^{2}-z_2+36z_1z_2-27
z_1z_2^{2},\\
dis(\hat{\bf K}_{{\bf F}_{2}})&=&(1-4z_1)^{2}-64z_1^{2}
z_2.
\label{dis}
\end{eqnarray}
With these results, M.~Naka determined the B-model Yukawa couplings of 
$K_{F_{n}}$
with respect to $u=\log(z_{1}), v=\log(z_{2})$ 
variables by assuming compatibility 
with the 
instanton expansion given by the double-log solution, and that they 
should 
be 
written in terms of simple rational functions multiplied by 
$1/dis(\hat{\bf K}_{{\bf F}_{n}})$ \cite{jn}:  
\\
\\
$n=0:$
\begin{eqnarray}
&&Y_{uuu}=\frac{-4z_1^{2}+4z_2^{2}-4z_1-2z_2+\frac{1}{4}}
{dis(\hat{\bf K}_{{\bf F}_{0}})},\;\;\;
Y_{uuv}=\frac{4z_1^{2}-4z_2^{2}+2z_2-\frac{1}{4}}
{dis(\hat{\bf K}_{{\bf F}_{0}})},\no\\
&&Y_{uvv}=\frac{-4z_1^{2}+4z_2^{2}+2z_1-\frac{1}{4}}
{dis(\hat{\bf K}_{{\bf F}_{0}})},\;\;\;
Y_{vvv}=\frac{4z_1^{2}-4z_2^{2}-2z_1-4z_2+\frac{1}{4}}
{dis(\hat{\bf K}_{{\bf F}_{0}})},\no
\end{eqnarray}
$n=1:$
\begin{eqnarray}
&&Y_{uuu}=\no\\
&&-\frac{((-162x+9)z_1z_2^{2}+(96x-4)z_1^{2}+(216x-14)z_1z_2+
(5-48x)z_1-6x(z_2-1))}
{dis(\hat{\bf K}_{{\bf F}_{1}})},\no\\
&&Y_{uuv}=\no\\
&&-\frac{((324x-18)z_1z_2^{2}+(8-192x)z_1^{2}+(25-432x)z_1z_2+
(96x-6)z_1+(12x-1)(z_2-1))}
{dis(\hat{\bf K}_{{\bf F}_{1}})},\no\\
&&Y_{uvv}=\no\\
&&-\frac{((36-648x)z_1z_2^{2}+(384x-16)z_1^{2}+(864x-44)z_1z_2+
(8-192x)z_1-(24x-1)(z_2-1))}
{dis(\hat{\bf K}_{{\bf F}_{1}})},\no\\
&&Y_{vvv}=\no\\
&&-\frac{((1296x-72)z_1z_2^{2}+(32-768x)z_1^{2}+(76-1728x)z_1z_2+
(384x-16)z_1+(48x-2)(z_2-1)-z_2)}
{dis(\hat{\bf K}_{{\bf F}_{1}})},
\no
\end{eqnarray}
$n=2:$
\begin{eqnarray}
&&Y_{uuu}=\frac{-1}{dis(\hat{\bf K}_{{\bf F}_{2}})},
\;\;\;Y_{uuv}=\frac{2z_1-\frac{1}{2}}
{dis(\hat{\bf K}_{{\bf F}_{2}})},\no\\
&&Y_{uvv}=\frac{-z_2(8z_1-1)}
{dis(\hat{\bf K}_{{\bf F}_{2}})(1-4z_2)},\;\;\;
Y_{vvv}=\frac{-z_2(24z_1z_2+2z_1-2z_2-\frac{1}{2})}
{dis(\hat{\bf K}_{{\bf F}_{2}})(1-4z_2)^2}.
\label{naka}
\end{eqnarray}
In the $n=1$ case, 
there exists a moduli parameter $x$ that leaves the instanton 
part of $Y_{ijk}$ invariant.\footnote{In the $n=0$ case, we also 
have one moduli parameter if we don't assume symmetry between $u$
and $v$.} 
In other words, we cannot determine 
the value of $x$ from the compatibility of the instanton numbers.
The aim of this section is to give a derivation 
of these Yukawa couplings by using an extended set 
Picard-Fuchs operators of local 
$F_{n}$ as the starting point. 
First, we notice that (\ref{naka}) tells 
us of the existence of a natural classical triple intersection theory 
on 
${K}_{F_{n}}$ 
compatible with the instanton expansion:\\
\\
$n=0:$
\begin{eqnarray}
&&\langle k_{u}k_{u}k_{u}\rangle
=\frac{1}{4},\;\;
\langle k_{u}k_{u}k_{v}\rangle 
=-\frac{1}{4},\;\;
\langle k_{u}k_{v}k_{v}\rangle 
=-\frac{1}{4},
\;\;\langle k_{v}k_{v}k_{v}\rangle
=\frac{1}{4},\no
\end{eqnarray}
$n=1:$
\begin{eqnarray}
&& \langle k_{u}k_{u}k_{u}\rangle  
=-6x,\;\;
\langle k_{u}k_{u}k_{v}\rangle 
=-1+12x,\;\;\langle k_{u}k_{v}k_{v}\rangle 
=-24x+1,\;\;
\langle k_{v}k_{v}k_{v}\rangle 
=-2+48x,\no
\end{eqnarray}
$n=2:$
\begin{eqnarray}
&&\langle k_{u}k_{u}k_{u}\rangle=-1,\;\;
\langle k_{u}k_{u}k_{v}\rangle
=-\frac{1}{2},\;\;
\langle k_{u}k_{v}k_{v}\rangle=0,\;\;
\langle k_{v}k_{v}k_{v}\rangle =0.
\label{cl}
\end{eqnarray}
In (\ref{cl}), we denote the classical triple intersection numbers of 
K\"ahler forms $k_{u},\;k_{v}$ by 
$\langle k_{u}k_{u}k_{u}\rangle,$ etc. 
Therefore, we have to reproduce (\ref{cl}) from the information 
obtained 
from some extended PF system. The key idea of constructing such an 
extended  
system becomes more clear upon looking at the triple log series 
obtained from 
 the generating hypergeometric series of
the solution of the PF system: 
\begin{equation}
w(u,v,r_{1},r_{2}):=
\sum_{n=0}^{\infty}\sum_{m=0}^{\infty}\frac{1}
{\prod_{j=0}^{4}\Gamma(1+l_{j}^{1}(m+r_1)+l_{j}^{2}(n+r_2))}
\exp((m+r_1)u+(n+r_2)v).
\label{gene}
\end{equation}
It is well-known that $w(u,v,r_{1},r_{2})|_{(r1,r2)=(0,0)}=1$ is 
the trivial solution of the PF system, and that the log solutions 
$\d_{r_{i}}w(u,v,r_{1},r_{2})|_{(r1,r2)=(0,0)}$ gives us the mirror map
$t^{i}$. Then we consider the relation between the triple log series 
$W_{ijk}(u,v):=
\d_{r_{i}}\d_{r_{j}}\d_{r_{k}}w(u,v,r_{1},r_{2})|_{(r1,r2)=(0,0)}$ 
and the prepotential $\mathcal F(t^{1},t^{2})$ of 
${K}_{F_{n}}$. Surprisingly, 
the classical intersection number (\ref{cl}) is determined from the
following assumption: 
\begin{eqnarray}
&&\frac{1}{6}\langle k_{u} k_{u}k_{u} \rangle
\cdot W_{111}(u(t_{*}),v(t_{*}))+
\frac{1}{2}\langle k_{u}k_{u}k_{v}\rangle
 \cdot W_{112}(u(t_{*}),v(t_{*}))\no\\
&&+\frac{1}{2}\langle k_{u}k_{v}k_{v}\rangle
\cdot W_{122}(u(t_{*}),v(t_{*}))+
\frac{1}{6}\langle k_{v}k_{v}k_{v}\rangle
\cdot W_{222}(u(t_{*}),v(t_{*}))\no\\
&&=t_{1}\frac{\d\mathcal F(t_{*})}{\d t_{1}}+t_{2}\frac{\d
\mathcal F(t_{*})}{\d t_{2}}-2\mathcal F(t_{*}).
\label{tri}
\end{eqnarray}
In the $n=0,2$ cases, we can determine classical intersection number
(\ref{cl}) uniquely from the instanton expansion of the r.h.s. of 
(\ref{tri}),
but in the $n=1$ case, we have one moduli parameter $x$ that leaves the 
instanton expansion invariant. Therefore, this situation is the same as 
in
Naka's result. With these intersection numbers, we can construct the 
set of 
relations of the classical cohomology ring of ``compactified''  
${K}_{F_{n}}$ (which we denote by $\overline{K}_{F_{n}}$) as 
follows:
\begin{eqnarray}
n=0:&&\no\\
&&k_u^{2}+k_v^{2}=0,\;\;k_uk_v^{2}-k_u^{2}k_v=0,\no\\
n=1:&&\no\\
&&(24x-1)(k_u^2-k_uk_v)-(18x-1)k_v^2=0,\;\;2k_uk_v^2+k_v^3=0, 
\;\;(x\neq\frac{1}{24})\no\\
&&k_v^2=0,\;\;2k_u^3-k_u^2k_v=0, \;\;(x=\frac{1}{24})\no\\
n=2:&&\no\\
&&k_v^2=0,\;\;k_u^{3}-2k_u^2k_v=0.
\label{clrel}
\end{eqnarray}
With this set up, we can construct an extended Picard-Fuchs system of
$\overline{K}_{F_{n}}$ that has the same principle part 
as (\ref{clrel}), and is constructed from the linear combination of 
$\mathcal D_{i},\;\;\theta_{1}\mathcal D_{i},\;\theta_{2}\mathcal 
D_{i}:$
$ (i=1,2)$:  
\begin{eqnarray}
&&n=0:\no\\
&&\tilde{\mathcal D}_{1}^{0}=
\mathcal D_{1}^{0}+\mathcal D_{2}^{0},\;\;\ 
\tilde{\mathcal D}_{2}^{0}=\theta_{1}
\mathcal D_{2}^{0}-\theta_{2}\mathcal D_{1}^{0},\no\\
&&n=1:\no\\
&& \tilde{\mathcal D}_{1}^{1}=
(24x-1)\mathcal D_{1}^{1}-(18x-1)\mathcal D_{2}^{1},
\;\;\tilde{\mathcal D}_{2}^{1}= 
(2\theta_{1}+\theta_{2})\mathcal D_{2}^{1},
\;\;(x\neq\frac{1}{24})\no\\
&&\tilde{\mathcal D}_{1}^{1}=
\mathcal D_{2}^{1},\;\;
\tilde{\mathcal D}_{2}^{1}=
(2\theta_{1}+\theta_{2})\mathcal 
D_{1}^{1},
\;\;(x=\frac{1}{24})\no\\
&&n=2:\no\\
&&
\tilde{\mathcal D}_{1}^{2}=
\mathcal D_{2}^{2},\;\;
\tilde{\mathcal D}_{2}^{2}=\theta_{1}\mathcal D_{1}^{2}. 
\label{expf}
\end{eqnarray}
In the remaining part of this section, 
we briefly discuss the derivation of the Yukawa coupling of 
$F_{0}$ in (\ref{naka}) by using (\ref{expf}) as the starting point. 
The other cases can be done in exactly the same way as in this 
computation. First, we use the standard definition of the B-model 
Yukawa 
coupling 
of mirror symmetry for a compact Calabi-Yau 3-fold: 
\begin{equation}
Y_{ijk}=\int_{\overline{\bf 
K}_{F_{0}}}\Omega\wedge\d_{i}\d_{j}\d_{k}\Omega.
\end{equation}
In the case of $\overline{K}_{F_{n}}$, the existence of a global 
holomomorphic 
three form $\Omega$ is not guaranteed, but we proceed here by assuming 
the 
existence 
of such an $\Omega$. 
We also apply the standard results obtained from Kodaira-Spencer 
theory 
on a compact Calabi-Yau 3-fold to the computation on 
$\overline{K}_{F_{0}}$. It is easy to show the following formula by 
application of this machinery: 
\begin{equation}
\int_{M}\Omega\wedge\d_{i}\d_{j}\d_{k}\d_{l}\Omega=
\frac{1}{2}(\d_{i}Y_{jkl}+\d_{j}Y_{ikl}+\d_{k}Y_{ijl}+\d_{l}Y_{ijk}).
\label{4}
\end{equation}
Next, we derive two relations among different Yukawa couplings obtained 
from 
$\theta_{1}(\mathcal D_{1}^{0}+\mathcal 
D_{2}^{0})\Omega=\theta_{2}(\mathcal 
D_{1}^{0}+\mathcal D_{2}^{0})\Omega=0$:
\begin{eqnarray}
&&Y_{uuu}+Y_{uvv}-4(z_1+z_2)(Y_{uuu}+2Y_{uuv}+Y_{uvv})=0,\no\\
&&Y_{uuv}+Y_{vvv}-4(z_1+z_2)(Y_{uuv}+2Y_{uvv}+Y_{vvv})=0.
\label{rly1}
\end{eqnarray}
We can also derive another relation from $(\theta_{1}
\mathcal D_{2}^{0}-\theta_{2}\mathcal D_{1}^{0})\Omega=0$:
\begin{equation}
Y_{uuv}-4z_1(Y_{uuv}+2Y_{uvv}+Y_{vvv})=
Y_{uvv}-4z_2(Y_{uuu}+2Y_{uuv}+Y_{uvv}).
\label{rly2}
\end{equation}
Since (\ref{rly1}) and (\ref{rly2}) are linear relations, we can easily 
solve 
them and obtain,
\begin{eqnarray}
&&Y_{uuu}=(1-16z_1-8z_2-16z_1^2+16z_2^2)S(u,v),\no\\
&&Y_{uuv}=(-1+8z_2+16z_1^2-16z_2^2)S(u,v),\no\\
&&Y_{uuv}=(-1+8z_1+16z_2^2-16z_1^2)S(u,v),\no\\
&&Y_{vvv}=(1-16z_2-8z_1-16z_2^2+16z_1^2)S(u,v).
\label{S}
\end{eqnarray} 
where $S(u,v)$ is an unknown function at this stage. Then we can derive 
differential equations of $S(u,v)$ from the relations 
$((\theta_{1})^{2}-(\theta_{2})^2)\mathcal D_{2}^{0}
\Omega=((\theta_{1})^{2}-(\theta_{2})^2)\mathcal D_{1}^{0}\Omega=0$ 
and (\ref{4}) by 
substituting the r.h.s. of (\ref{S}). These operations result in 
the following differential equations of $S(u,v)$:
\begin{eqnarray}
&&\frac{\d_{u}S(u,v)}{S(u,v)}=\frac{8z_1-32z_1^2+32z_1z_2}
{1-8z_1-8z_2+16z_1^2-32z_1z_2+16z_2^2},\no\\
&&\frac{\d_{v}S(u,v)}{S(u,v)}=\frac{8z_2-32z_2^2+32z_1z_2}
{1-8z_1-8z_2+16z_1^2-32z_1z_2+16z_2^2}.
\end{eqnarray}
We can immediately solve the above equations and obtain,
\begin{equation}
S(u,v)=(const.)\cdot\frac{1}{1-8z_1-8z_2+16z_1^2-32z_1z_2+16z_2^2}.
\end{equation}
Finally, the classical intersection numbers in (\ref{cl}) tell us that 
$(const.)=\frac{1}{4}$.

\section{The del Pezzo surface $K_{dP_2}$.}

We can also look to a three parameter model, in order to determine what 
we might expect in more general situations. The examples of the two 
parameter case might cause one to hope that, for every local geometry 
of 
the form $K_S$, we can extend the original PF system to give a complete 
description of mirror symmetry from the $B$ model geometry alone. Here 
we will demonstrate that this is indeed the case for $K_{dP_2}$. 
However, 
it is no longer necessary to use a higher order system for three and 
higher parameter cases; we can find a complete set of solutions by 
``forgetting" about some of the originally proposed local mirror 
symmetry 
operators. 

The symplectic quotient description of $K_{dP_2}$ may be written as
$$
\{(w_1,\dots,w_6)\in \C^6-Z: \sum_{k=1}^6 l_k^i |w_k|^2=r_i, \ 
i=1,2,3\}/(S^1)^3
$$
with the vectors 
$$
\begin{pmatrix}
 l^1 \\ l^2 \\ l^3  
\end{pmatrix}=
\begin{pmatrix}
	-1 & 1 & -1 & 1 & 0 & 0 \\ -1 & -1 & 1 & 0 & 0 & 1 \\ -1 & 0 &1&-1&1&0
 \end{pmatrix}.
$$
Note that $b_2=3$ and $b_4=1$.

\begin{figure}[htbp]
\centering
\input{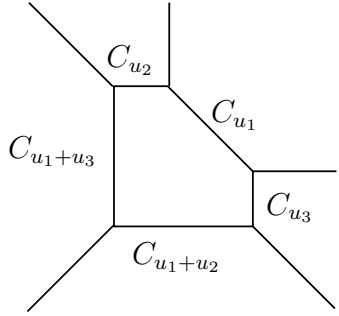}
\caption{The del Pezzo $dP_2$. Each curve $C_{u_i}$ corresponds to a 
vector $l^i$.}
\label{delpezzo}
\end{figure}

This comes with PF operators 
\begin{eqnarray}
\mathcal D_1&=& 
(\theta_1-\theta_2)(\theta_1-\theta_3)-z_1(\theta_1+\theta_2+\theta_3)(\theta_1-\theta_2-\theta_3),\no\\
\mathcal D_2&=& 
(\theta_2+\theta_3-\theta_1)\theta_2-z_2(\theta_1+\theta_2+\theta_3)(\theta_2-\theta_1),\no\\
\mathcal D_3&=& 
(\theta_2+\theta_3-\theta_1)\theta_3-z_3(\theta_1+\theta_2+\theta_3)(\theta_3-\theta_1),\no\\
\mathcal D_{4}&=&\theta_{2}(\theta_{1}-\theta_{3})-z_{1}z_{2}
(\theta_{1}+\theta_{2}+\theta_{3}+1)(\theta_{1}+\theta_{2}+\theta_{3}),\no\\
\mathcal D_{5}&=&\theta_{3}(\theta_{1}-\theta_{2})-z_{1}z_{3}
(\theta_{1}+\theta_{2}+\theta_{3}+1)(\theta_{1}+\theta_{2}+\theta_{3}).
\label{pfdp2}
\end{eqnarray}
Let $\mathcal F$, $t_1,t_2,t_3$ and the $\Pi_{ij}$'s be as before. Then 
\cite{CKYZ} provides a single double logarithmic solution $W$, 
corresponding to the four cycle in the base of the $A$ model geometry, 
which 
satisfies
$$
W=\Pi_{11}+\Pi_{12}+\Pi_{13}+\Pi_{23}=\frac{\partial 
\mathcal F}{\partial t_1}+\frac{\partial \mathcal F}{\partial 
t_2}+\frac{\partial \mathcal F}{\partial t_3}.
$$ 

The naive approach in this case suggests that we need to add two 4 
cycles to the $A$ model geometry. Using as motivation the notion that 
each 
double logarithmic solution of the system should correspond to a 4 
cycle on the mirror, we are led to work with a system consisting of 3 
of 
the original 5 PF differential operators. For reasons to be discussed 
shortly, we will use the set
\begin{eqnarray}
\tilde{\mathcal D}_1&=&(6x+2y-1)(\mathcal D_1-\mathcal D_2-\mathcal D_3) -x(\mathcal D_{2}+\mathcal 
D_{3}),\no\\
\tilde{\mathcal D}_2&=&(5x+y-1)(\mathcal D_2+\mathcal D_3+\mathcal D_4+\mathcal D_5)+
(x+y)(\mathcal D_4+\mathcal D_5),\no\\
\tilde{\mathcal D}_3&=&(x-y-1)(\mathcal D_2-\mathcal D_3+\mathcal D_4-\mathcal D_5)+(x-y)(\mathcal D_{4}-\mathcal D_5),
\label{epfdp2}
\end{eqnarray}
where $x$ and $y$ are real free parameters. 
If we set $(x,y)=(0,0)$, we can easily show that a basis of the solution space with double 
logarithmic singularities is three dimensional, and is provided by
\begin{eqnarray}
\Pi_{12}=\frac{\partial \mathcal F}{\partial t_2},\no\\
\Pi_{13}=\frac{\partial \mathcal F}{\partial t_3}, \no\\
\Pi_{11}+\Pi_{23}=\frac{\partial \mathcal F}{\partial t_1}.
\end{eqnarray}

Hence, we can recover the full prepotential from solutions of the new 
PF system alone. We expect that this phenomenon will continue to hold 
true in all cases of the type $K_S$. 

The extended PF system of $K_{dP_{2}}$ (\ref{epfdp2}) 
is derived in the same way as $K_{F_{n}}$. Let us introduce the 
logarithm 
of 
the standard B-model coordinates $z_{i}$:
\begin{equation}
u_{1}=\log(z_{1}),\;\;u_{2}=\log(z_{2}),\;\;u_{3}=\log(z_{3}),
\end{equation}
and consider the generating function of solutions of the PF system 
(\ref{pfdp2}): 
\begin{eqnarray}
w(u_{1},u_{2},u_{3},r_{1},r_{2},r_{3}):=
&&\sum_{n_{1}=0}^{\infty}\sum_{n_{2}=0}^{\infty}\sum_{n_{3}=0}^{\infty}\frac{1}
{\prod_{j=0}^{5}
\Gamma(1+l_j^{1}(n_{1}+r_1)+l_j^{2}(n_{2}+r_2)+
l_{j}^{3}(n_{3}+r_3))}\times\no\\
&&\times\exp((n_{1}+r_1)u_{1}+(n_{2}+r_2)u_{2}+(n_{3}+r_3)u_{3}).
\label{genedP}
\end{eqnarray}
The mirror map of $K_{dP_{2}}$ is given by  
\begin{equation}
t_{i}(u_{*}):=\d_{r_{i}}w(u_{1},u_{2},u_{3},r_{1},r_{2},r_{3})
|_{r_{i}=0}.
\end{equation}
From the triple log function
\begin{equation}
W_{ijk}(u_{1},u_{2},u_{3})
:=\d_{r_{i}}\d_{r_{2}}\d_{r_{3}}w(u_{1},u_{2},u_{3},r_{1},r_{2},r_{3})
|_{r_{i}=0},
\end{equation}
obtained from (\ref{genedP}), we can find the classical triple 
intersection 
number $ \langle k_{u_{i}}k_{u_{j}}k_{u_{m}}\rangle $ of K\"ahler forms 
$k_{u_{i}}$ by assuming the following relation: 
\begin{eqnarray}
&&\frac{1}{6}\langle k_{u_{1}}k_{u_{1}}k_{u_{1}}\rangle
\cdot 
W_{111}(u_{1}(t_{*}),u_{2}(t_{*}),u_{3}(t_{*}))+
\frac{1}{6}\langle  k_{u_{2}}k_{u_{2}}k_{u_{2}}\rangle 
\cdot W_{222}(u_{1}(t_{*}),
u_{2}(t_{*}),u_{3}(t_{*}))+\no\\
&&\frac{1}{6}\langle  k_{u_{3}}k_{u_{3}}k_{u_{3}}\rangle 
\cdot W_{333}(u_{1}(t_{*}),
u_{2}(t_{*}),u_{3}(t_{*}))+\no\\
&&\frac{1}{2}\langle  k_{u_{1}}k_{u_{1}}k_{u_{2}}\rangle
\cdot W_{112}(u_{1}(t_{*}),
u_{2}(t_{*}),u_{3}(t_{*}))+
\frac{1}{2}\langle  k_{u_{1}}k_{u_{2}}k_{u_{2}}\rangle 
\cdot W_{122}(u_{1}(t_{*}),
u_{2}(t_{*}),u_{3}(t_{*}))+\no\\
&&\frac{1}{2}\langle  k_{u_{2}}k_{u_{2}}k_{u_{3}} \rangle
\cdot W_{223}(u_{1}(t_{*}),
u_{2}(t_{*}),u_{3}(t_{*}))+
\frac{1}{2}\langle  k_{u_{2}}k_{u_{3}}k_{u_{3}}\rangle 
\cdot W_{233}(u_{1}(t_{*}),
u_{2}(t_{*}),u_{3}(t_{*}))+\no\\
&&\frac{1}{2}\langle  k_{u_{1}}k_{u_{1}}k_{u_{3}}\rangle
\cdot W_{113}(u_{1}(t_{*}),
u_{2}(t_{*}),u_{3}(t_{*}))+
\frac{1}{2}\langle  k_{u_{1}}k_{u_{3}}k_{u_{3}}\rangle
\cdot  W_{133}(u_{1}(t_{*}),
u_{2}(t_{*}),u_{3}(t_{*}))+\no\\
&& \langle  k_{u_{1}}k_{u_{2}}k_{u_{3}}\rangle 
\cdot W_{123}(u_{1}(t_{*}),
u_{2}(t_{*}),u_{3}(t_{*}))\no\\
&&=t_{1}\frac{\d\mathcal F(t_{*})}{\d t_{1}}+t_{2}
\frac{\d\mathcal F(t_{*})}{\d t_{2}}+
t_{3}\frac{\d\mathcal F(t_{*})}{\d t_{3}}-2\mathcal 
F(t_{*}),
\label{rdp}
\end{eqnarray}
where we used instanton expansion part of $\mathcal 
F(t_{1},t_{2},t_{3})$
read off from double log solution of (\ref{pfdp2}).
Taking the symmetry between $k_{u_{2}}$ and $k_{u_{3}}$ into account, 
we found the following classical intersection numbers with two free parameters\footnote
{In this case,  we have four moduli parameters unless we assume symmetry 
between $k_{u_{2}}$ and $k_{u_{3}}$.} $x$ and $y$:
\begin{eqnarray}
&&\langle  k_{u_{1}}k_{u_{1}}k_{u_{1}}\rangle=
-1+6x+2y,\;\;\;
\langle  k_{u_{2}}k_{u_{2}}k_{u_{2}}\rangle=-y,\;\;\;
\langle k_{u_{3}}k_{u_{3}}k_{u_{3}}\rangle=-y,\;\;\;
\langle k_{u_{1}}k_{u_{1}}k_{u_{2}}\rangle=-3x-y,\;\;\;
\no\\&&
\langle k_{u_{1}}k_{u_{2}}k_{u_{2}}\rangle=x+y,\;\;\;
\langle k_{u_{2}}k_{u_{2}}k_{u_{3}}\rangle=-x,\;\;\;
\langle k_{u_{2}}k_{u_{3}}k_{u_{3}}\rangle=-x,\;\;\;
\langle k_{u_{1}}k_{u_{1}}k_{u_{3}}\rangle=-3x-y,\;\;\;
\no\\&&
\langle k_{u_{1}}k_{u_{3}}k_{u_{3}}\rangle=x+y,\;\;\;
\langle k_{u_{1}}k_{u_{2}}k_{u_{3}}\rangle=-1+2x.
\label{cldp}
\end{eqnarray}
With this data, we can construct a complete set of relations of 
$k_{u^{i}}$ 
that reproduce (\ref{cldp}) as follows:
\begin{eqnarray}
R_1&=&(6x+2y-1)(p_1-p_2-p_3) -x(p_{2}+p_{3}),\no\\
R_2&=&(5x+y-1)(p_2+p_3+p_4+p_5)+
(x+y)(p_4+p_5),\no\\
R_3&=&(x-y-1)(p_2-p_3+p_4-p_5)+(x-y)(p_4-p_5),
\label{ecl}
\end{eqnarray}
where
\begin{eqnarray}
&&p_1=(k_{u_1}-k_{u_2})(k_{u_1}-k_{u_3}),\;p_2=k_{u_2}(k_{u_2}+k_{u_3}-k_{u_1}),\;
p_3=k_{u_3}(k_{u_2}+k_{u_3}-k_{u_1}),\no\\
&&p_4=k_{u_2}(k_{u_1}-k_{u_3}),\;p_5=k_{u_3}(k_{u_1}-k_{u_2}).
\end{eqnarray}  
The extended PF system (\ref{epfdp2}) is obtained from linear 
combinations 
of the $\mathcal D_{i}$'s that reduce to (\ref{ecl}) at the large 
radius 
limit. Of course, we can compute the B-model Yukawa coupling of 
$K_{dP_{2}}$ 
by using 
(\ref{epfdp2}) as the starting point in the same way as $F_{0}$, and we 
collect the resulting Yukawa couplings in Appendix B.  

\section{Conclusion.}

Through a variety of examples, we have seen the emergence of a new set 
of differential operators for local mirror symmetry. In this sense, we 
may view this as a next step towards a complete treatment of the 
program initiated in \cite{CKYZ}.  However, at present, we lack an 
understanding of how to choose classical intersection theory.  Is 
there, in fact, 
a canonical way to associate triple intersection numbers on $K_S$, as 
we saw in the $\dim H_4(X,\Z)=0$ cases? Also, although the methods 
presented here produce results consistent with physical expectations, 
it is 
unsatisfying to be without a better geometric understanding of the 
extended PF systems. It would be interesting to find compact spaces 
whose 
period integrals agree with the solutions of the extended PF systems. 
We 
leave these questions for future work.

\newpage

\section*{Appendix A : An extended Picard-Fuchs system for $X_{3}$.}

\begin{figure}[htbp]
\centering
\input{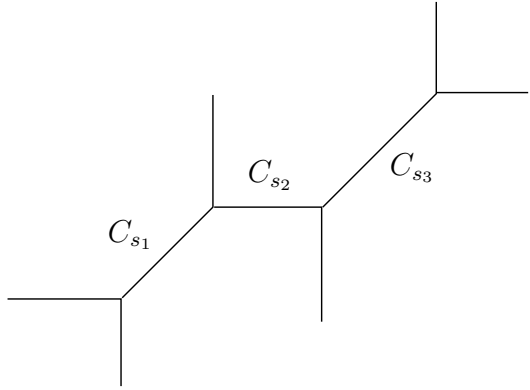}
\caption{Toric graph for $X_3$.}
\label{threeminusones}
\end{figure}

Here, we apply our conjecture on intersection theory for $X$ satisfying 
$\dim H_4(X,\Z)=0$ to a three parameter example, in order to more fully 
explore its applicability. We will work with $X_3=$ 
$$
\{(w_1,\dots,w_6)\in \C^6-Z: \sum_{k=1}^6 l_k^i |w_k|^2=r_i, \ 
i=1,2,3\}/(S^1)^3
$$
where
\begin{equation}
\begin{pmatrix}
 l^1 \\ l^2 \\ l^3  
\end{pmatrix}=
\begin{pmatrix}
	1 & 1 & -1 & -1 & 0 & 0 \\ 0 & -1 & -1 & 1 & 1 & 0 \\ 0 & -1 &1&0&-1&1
 \end{pmatrix}.
\end{equation}
The toric graph is provided in figure \ref{threeminusones}. Then $X_3$ has 
no 4 cycle, and for each curve $C_{s_i}$ corresponding to the secondary 
fan vector $l^i$, we have that $\mathcal N_{C_{s_i/X_3}}\cong \mathcal 
O (-1) \oplus \mathcal O (-1)$.
\noindent
By making use of the instanton parts given in \cite{I} and our 
conjecture, we immediately have the Yukawa couplings: \newline
{\bf A-model Yukawa Couplings of $X_{3}$ (w.r.t. $s_{i}$) }
\begin{eqnarray}
Y_{111}&=&\frac{1}{2}+\frac{e^{s_{1}}}{1-e^{s_{1}}}-
\frac{e^{s_{1}+s_{2}}}{1-e^{s_{1}+s_{2}}}+
\frac{e^{s_{1}+s_{2}+s_{3}}}{1-e^{s_{1}+s_{2}+s_{3}}},
\no\\
Y_{113}&=&\frac{1}{2}+
\frac{e^{s_{1}+s_{2}+s_{3}}}{1-e^{s_{1}+s_{2}+s_{3}}},
\no\\
Y_{113}&=&\frac{1}{2}+
\frac{e^{s_{1}+s_{2}+s_{3}}}{1-e^{s_{1}+s_{2}+s_{3}}},
\no\\
Y_{333}&=&\frac{1}{2}+\frac{e^{s_{3}}}{1-e^{s_{3}}}-
\frac{e^{s_{3}+s_{2}}}{1-e^{s_{3}+s_{2}}}+
\frac{e^{s_{1}+s_{2}+s_{3}}}{1-e^{s_{1}+s_{2}+s_{3}}},
\no
\end{eqnarray}
\begin{eqnarray}
Y_{112}&=&-\frac{e^{s_{1}+s_{2}}}{1-e^{s_{1}+s_{2}}}+
\frac{e^{s_{1}+s_{2}+s_{3}}}{1-e^{s_{1}+s_{2}+s_{3}}},
\no\\
Y_{233}&=&-\frac{e^{s_{3}+s_{2}}}{1-e^{s_{3}+s_{2}}}+
\frac{e^{s_{1}+s_{2}+s_{3}}}{1-e^{s_{1}+s_{2}+s_{3}}},
\no\\
Y_{122}&=&-\frac{e^{s_{1}+s_{2}}}{1-e^{s_{1}+s_{2}}}+
\frac{e^{s_{1}+s_{2}+s_{3}}}{1-e^{s_{1}+s_{2}+s_{3}}},
\no\\
Y_{223}&=&-\frac{e^{s_{3}+s_{2}}}{1-e^{s_{3}+s_{2}}}+
\frac{e^{s_{1}+s_{2}+s_{3}}}{1-e^{s_{1}+s_{2}+s_{3}}},
\no\\
Y_{222}&=&\frac{e^{s_{2}}}{1-e^{s_{2}}}-
\frac{e^{s_{1}+s_{2}}}{1-e^{s_{1}+s_{2}}}-
\frac{e^{s_{3}+s_{2}}}{1-e^{s_{3}+s_{2}}}+
\frac{e^{s_{1}+s_{2}+s_{3}}}{1-e^{s_{1}+s_{2}+s_{3}}},
\no\\
Y_{123}&=&\frac{1}{2}+
\frac{e^{s_{1}+s_{2}+s_{3}}}{1-e^{s_{1}+s_{2}+s_{3}}}.
\end{eqnarray}
Once we have these, the rest of the calculations are routine. We list 
the results here. \newline
{\bf Ordinary  Picard-Fuchs system of $X_{3}$ }
\begin{eqnarray}
&&\mathcal D_{1}=\theta_{1}(\theta_{1}-\theta_{2}-\theta_{3})-
z_{1}(\theta_{1}-\theta_{2})(\theta_{1}+\theta_{2}-\theta_{3}),
\no\\
&&\mathcal D_{2}=(\theta_{2}-\theta_{1})(\theta_{2}-\theta_{3})-
z_{2}(\theta_{2}+\theta_{3}-\theta_{1})(\theta_{2}+\theta_{1}-\theta_{3}),
\no\\&&
\mathcal D_{3}=\theta_{3}(\theta_{3}-\theta_{2}-\theta_{1})-
z_{3}(\theta_{3}-\theta_{2})(\theta_{3}+\theta_{2}-\theta_{1}),
\no\\&&
\mathcal D_{4}=\theta_{1}(\theta_{1}-\theta_{3})-
z_{1}z_{2}(\theta_{1}+\theta_{2}-\theta_{3}+1)
(\theta_{1}+\theta_{2}-\theta_{3}),
\no\\&&
\mathcal D_{5}=\theta_{1}\theta_{3}-
z_{1}z_{3}(\theta_{2}-\theta_{1})(\theta_{2}-\theta_{3}),
\no\\&&
\mathcal D_{6}=\theta_{3}(\theta_{3}-\theta_{1})-
z_{2}z_{3}(\theta_{3}+\theta_{2}-\theta_{1}+1)
(\theta_{3}+\theta_{2}-\theta_{1}).
\label{pfx3}
\end{eqnarray}
{\bf Jacobian of the Mirror Map}
\begin{eqnarray}
&&u_{1}=\log(z_{1}),\;\;u_{2}=\log(z_{2}),\;\;u_{3}=\log(z_{3}).\no\\
&&\frac{\d s_1}{\d u_1}=\frac{1}{2}
\frac{-\sqrt{1-4z_1z_2}-1+4z_1z_2}{4z_1z_2-1},\no\\
&&\frac{\d s_1}{\d u_2}=\frac{1}{2}
\frac{4z_1z_2\sqrt{1-4z_2z_3}+\sqrt{1-4z_1z_2}-\sqrt{1-4z_2z_3}-
4z_2z_3\sqrt{1-4z_1z_2}}{(4z_1z_2-1)(4z_2z_3-1)},\no\\
&&\frac{\d s_1}{\d u_3}=\frac{1}{2}
\frac{-1+4z_2z_3+\sqrt{1-4z_2z_3}}{4z_2z_3-1},\no\\
&&\frac{\d s_2}{\d u_1}=-\frac{1}{2}
\frac{-1+4z_1z_2+\sqrt{1-4z_1z_2}}{4z_1z_2-1},\no
\end{eqnarray}
\begin{eqnarray}
&&\frac{\d s_2}{\d u_2}=-\frac{1}{2}
\frac{-\sqrt{1-4z_1z_2}+4z_1z_2\sqrt{1-4z_2z_3}-\sqrt{1-4z_2z_3}+
4z_2z_3\sqrt{1-4z_1z_2}}{(4z_1z_2-1)(4z_2z_3-1)},\no\\
&&\frac{\d s_2}{\d u_3}=-\frac{1}{2}
\frac{-1+4z_2z_3+\sqrt{1-4z_2z_3}}{4z_2z_3-1},\no\\
&&\frac{\d s_3}{\d u_1}=\frac{1}{2}
\frac{-1+4z_1z_2+\sqrt{1-4z_1z_2}}{4z_1z_2-1},\no\\
&&\frac{\d s_3}{\d u_2}=\frac{1}{2}
\frac{\sqrt{1-4z_2z_3}+4z_2z_3\sqrt{1-4z_1z_2}-4z_1z_2\sqrt{1-4z_2z_3}
-\sqrt{1-4z_1z_2}}{(4z_1z_2-1)(4z_2z_3-1)},\no\\
&&\frac{\d s_3}{\d u_3}=\frac{1}{2}
\frac{-\sqrt{1-4z_2z_3}-1+4z_2z_3}{4z_2z_3-1}.
\end{eqnarray}
{\bf B-model Yukawa Couplings of $X_{3}$ (w.r.t $u_{i}$)}
\begin{eqnarray}
Y_{111}&=& 
(8z_2^3z_1^4z_3^2-10z_3^2z_1^3z_2^2+3z_1^2z_2z_3^2+8z_3z_1^3z_2^2-
24z_3z_1^2z_2^2-2z_3z_2z_1^2+12z_3z_1z_2-z_3-
\no\\&&
z_3z_1-8z_1^2z_2^3+16z_1^2z_2^2+10z_1z_2^2-z_2-12z_1z_2+1+z_1)/(\Delta
(4z_1z_2-1)^2),\no\\
&&\no\\
Y_{113} &=& 
(2z_3^2z_1^3z_2^2-z_1^2z_2z_3^2+4z_3z_2z_1^2-4z_3z_1z_2-z_3z_1+z_3-
2z_1z_2^2-2z_2z_1^2+2z_1z_2+z_2+z_1-1)\no\\&&
/(\Delta(4z_1z_2-1)),\no\\
&&\no\\
Y_{133} &=& 
(2z_1^2z_2^2z_3^3-2z_3z_2^2-z_1^2z_2z_3^2+4z_3^2z_2z_1-4z_3z_1z_2+z_2+2z_2z_3-2z_3^2z_2-z_3z_1+z_1+z_3-1)\no\\&&
/(\Delta(4z_2z_3-1)),\no\\
&&\no\\
Y_{333} &=& 
(8z_3^4z_2^3z_1^2-8z_3^2z_2^3-10z_1^2z_2^2z_3^3+8z_3^3z_2^2z_1-24z_3^2z_1z_2^2+16z_3^2z_2^2+10z_3z_2^2+\no\\
&&
3z_1^2z_2z_3^2-2z_3^2z_2z_1+12z_3z_1z_2-z_2-12z_2z_3+1+z_3-z_1-z_3z_1)
/(\Delta(4z_2z_3-1)^2),\no
\end{eqnarray}
\begin{eqnarray}
Y_{112} &=& 
-2z_2z_1(2z_1^2z_2z_3^2-z_3^2z_1-8z_3z_2z_1^2+4z_3z_1z_2+4z_3z_1-2z_3+
4z_2z_1^2-2z_2-3z_1+2)\no\\&&
/(\Delta(4z_1z_2-1)^2),
\no\\
&&\no\\
Y_{233} &=& 
-2z_2z_3(2z_1^2z_2z_3^2-8z_3^2z_2z_1+4z_3^2z_2+4z_3z_1z_2+4z_3z_1-3z_3-z_3z_1^2-2z_2-2z_1+2)\no\\&&
/(\Delta(4z_2z_3-1)^2),\no
\end{eqnarray}
\begin{eqnarray}
Y_{122} &=& 
-2(4z_3^3z_2^2z_1^3-3z_3^3z_2z_1^2-16z_3^2z_1^3z_2^2+4z_3^2z_1^4z_2^2+12z_1^2z_2z_3^2-3z_3^2z_2z_1^3+\no\\&&
z_1^2z_3^2-z_3^2-4z_2^2z_3z_1+16z_3z_1^2z_2^2-z_2z_3+4z_3z_2z_1^3-12z_3z_2z_1^2+2z_3z_1+z_3-3z_3z_1^2-\no\\&&
2z_1+2z_1^2+3z_1z_2-4z_1^2z_2^2)z_2/(\Delta(4z_2z_3-1)(4z_1z_2-1)^2),\no
\end{eqnarray}
\begin{eqnarray}
Y_{223} 
&=&-2(4z_3^4z_2^2z_1^2+4z_3^3z_2^2z_1^3-16z_1^2z_2^2z_3^3-3z_3^3z_2z_1^2+4z_3^3z_2z_1+16z_3^2z_1z_2^2-4z_3^2z_2^2-
\no\\&&
3z_3^2z_2z_1^3+12z_1^2z_2z_3^2-12z_3^2z_2z_1+z_1^2z_3^2-3z_3^2z_1+2z_3^2-4z_2^2z_3z_1+3z_2z_3+2z_3z_1-2z_3-\no\\&&
z_1z_2-z_1^2+z_1)z_2/(\Delta(4z_2z_3-1)^2(4z_1z_2-1)),\no
\end{eqnarray}
\begin{eqnarray}
Y_{222} &=& 
2(1+32z_3z_1z_2-28z_3z_2z_1^2-32z_3z_1^2z_2^2+32z_1^2z_2z_3^2-32z_3^2z_1^3z_2^2-28z_3^2z_2z_1-4z_3^2z_2z_1^3-
\no\\&&
32z_3^2z_1z_2^2-4z_3^3z_2z_1^2-32z_1^2z_2^2z_3^3+8z_3^4z_2^2z_1^2+8z_3^2z_1^4z_2^2+96z_3^2z_2^2z_1^2-2z_1-2z_3+\no\\&&
8z_1^2z_2^2-4z_1z_2+4z_3z_1+8z_3^2z_2^2+z_1^2z_3^2-2z_3^2z_1-4z_2z_3+z_1^2+z_3^2-2z_3z_1^2+4z_1^3z_2+\no\\&&
4z_3^3z_2)z_2/(\Delta(4z_2z_3-1)^2(4z_1z_2-1)^2),\no
\end{eqnarray}
\begin{eqnarray}
Y_{123} &=& 
-(2z_1^2z_2^2z_3^3-16z_3^2z_2^2z_1^2+8z_3^2z_1z_2^2+2z_3^2z_1^3z_2^2+4z_3^2z_2z_1-z_1^2z_2z_3^2-2z_3^2z_2-2z_3z_2^2+
\no\\&&
8z_3z_1^2z_2^2-8z_3z_1z_2+4z_3z_2z_1^2+2z_2z_3-z_3z_1+z_3-1+
2z_1z_2-2z_1z_2^2+z_1-2z_2z_1^2z_2)/
\no\\&&
(\Delta(4z_2z_3-1)(4z_1z_2-1)).
\end{eqnarray}
{\bf Discriminant of $X_{3}$}
\begin{eqnarray}
\Delta=2(-z_1^2z_2z_3^2+2z_3z_1z_2+z_3z_1-z_3-z_2-z_1+1).
\end{eqnarray}
{\bf Extended Picard-Fuchs system of $X_{3}$}
\begin{eqnarray}
\tilde{\mathcal 
D}_{1}&=&a_{1}(z_{1},z_{2},z_{3})\cdot\mathcal 
D_{1}+a_{2}(z_{1},z_{2},z_{3})
\cdot\mathcal D_{2}+a_{3}(z_{1},z_{2},z_{3})\cdot\mathcal D_{3}+
a_{5}(z_{1},z_{2},z_{3})\cdot\mathcal D_{5}-\no\\
&&2(1+z_1+z_2+z_3-2z_1z_2z_3+z_1z_3+z_1^2z_2z_3^2)\mathcal D_{6}\no\\
&=&(-\theta_2^2-2\theta_2\theta_3-\theta_1^2+2\theta_2\theta_1+3\theta_3^2
)z_1^2z_2z_3^2+(((-4\theta_1^2+4\theta_2^2+8\theta_1\theta_3+
4\theta_2\theta_3-8\theta_3^2)z_1+\no\\
&&(4\theta_1\theta_3-4\theta_3^2+
4\theta_2\theta_3)z_1^2)z_2+(-2\theta_2\theta_3+2\theta_2\theta_1-
2\theta_1\theta_3-\theta_1^2-\theta_2^2+3\theta_3^2)z_1-\theta_2^2-
\no\\&&
2\theta_2\theta_3-2\theta_1\theta_3+3\theta_3^2+\theta_1^2)z_3+
((-4\theta_1^2+4\theta_2^2+4\theta_1\theta_3-4\theta_2\theta_3)z_1+
\theta_1^2+\theta_3^2-2\theta_2\theta_1+
\no\\&&
\theta_2^2-2\theta_2\theta_3)z_2+(-\theta_1^2+
\theta_3^2+2\theta_2\theta_1-\theta_2^2-2\theta_1\theta_3)z_1+
\theta_1^2+\theta_3^2-\theta_2^2-
2\theta_1\theta_3-\no\\
&&
2(1+z_1+z_2+z_3-2z_1z_2z_3+z_1z_3+z_1^2z_2z_3^2)\mathcal D_6,\no\\
\tilde{\mathcal D}_{2}&=&\mbox{interchange between $z_{1}$ and $z_{3}$ 
of 
$\tilde{\mathcal D}_{1}$},\no\\
\tilde{\mathcal D}_{3}&=&(1+z_3)\mathcal D_{1}+(1+z_1)\mathcal D_{3}+
(1-z_1^2z_3^2)\mathcal D_{2}+(1-z_1z_3)\mathcal D_{5},
\end{eqnarray}
where $a_{j}(z_{1},z_{2},z_{3})$ are rational functions in $z_{i}$ 
whose 
denominators are given by $\Delta$.
\section*{Appendix B : B-model Yukawa couplings of ${K}_{dP_2}$}
In this Appendix, we present the B-model Yukawa couplings of 
${K}_{dP_2}$
computed from the extended PF system (\ref{epfdp2}), and the assumption 
of the existence of Kodaira-Spencer 
theory:
\begin{equation}
Y_{ijk}=\int_{\overline{K}_{dP_{2}}}\Omega\wedge\frac{\d^{3}\Omega}
{\d u_{i}\d u_{j}\d {u_{k}}}.
\end{equation}
\begin{eqnarray}
Y_{222}&=&-(16z_3^3yz_1^2+((-27y-18)z_1z_2^2+16yz_1^3-16yz_1^2+\no\\&&
((-24y-8)z_1^2+(4+36y)z_1)z_2-8z_1y)z_3^2+(((-24y-14)z_1^2+\no\\&&
(22+36y)z_1)z_2^2-8yz_1^2+((64y+12)z_1^2+(-2-46y)z_1+(-32y-8)z_1^3-\no\\&&
y-1)z_2+8z_1y+y)z_3+(12+16y)z_1^2z_2^3+((16y+8)z_1^3+(-16y-4)z_1^2+\no\\&&
(-3-8y)z_1)z_2^2+(y+(2-8y)z_1^2+1+(-3+8y)z_1)z_2+z_1y-y)/\Delta,
\no\\
&&\no\\
Y_{223}&=&-(16z_3^3xz_1^2+((-27x+9)z_1z_2^2+16xz_1^3-16xz_1^2+((-24x+8)z_1^2+\no\\&&
(36x-8)z_1)z_2-8z_1x)z_3^2+(((36x-11)z_1+(-24x+6)z_1^2)z_2^2-8xz_1^2+\no\\&&
((-16+64x)z_1^2+(-32x+8)z_1^3+(10-46x)z_1-x)z_2+8z_1x+x)z_3+(16x-8)z_1^2z_2^3+\no\\&&
((-16x+8)z_1^2+(16x-8)z_1^3+(-8x+2)z_1)z_2^2+((8x-2)z_1+(-8x+2)z_1^2+x)z_2+\no\\&&
z_1x-x)/\Delta,
\no\\
&&\no\\
Y_{233}&=&-((16x-8)z_1^2z_3^3+((-27x+9)z_1z_2^2+(16x-8)z_1^3+(-16x+8)z_1^2+\no\\&&
((36x-11)z_1+(-24x+6)z_1^2)z_2+(-8x+2)z_1)z_3^2+(((-24x+8)z_1^2+\no\\&&
(36x-8)z_1)z_2^2+(-8x+2)z_1^2+((-16+64x)z_1^2+(-32x+8)z_1^3+(10-46x)z_1-x)z_2+\no\\&&
(8x-2)z_1+x)z_3+16z_2^3xz_1^2+(-16xz_1^2+16xz_1^3-8z_1x)z_2^2+\no\\&&
(8z_1x-8xz_1^2+x)z_2+z_1x-x)/\Delta,
\no\\
&&\no\\
Y_{333}&=&-((12+16y)z_1^2z_3^3+((-27y-18)z_1z_2^2+(-16y-4)z_1^2+((-24y-14)z_1^2+\no\\&&
(22+36y)z_1)z_2+(16y+8)z_1^3+(-3-8y)z_1)z_3^2+((2-8y)z_1^2+((-24y-8)z_1^2+\no\\&&
(4+36y)z_1)z_2^2+(-3+8y)z_1+1+((64y+12)z_1^2+(-2-46y)z_1+(-32y-8)z_1^3-\no\\&&
y-1)z_2+y)z_3+16z_2^3yz_1^2+(16yz_1^3-16yz_1^2-8z_1y)z_2^2+z_1y+\no\\&&
(-8yz_1^2+y+8z_1y)z_2-y)/\Delta,\no
\end{eqnarray}
\begin{eqnarray}
Y_{122} 
&=& ((16x+16y)z_1^2z_3^3+((16x+16y)z_1^3+(-27y-9-27x)z_1z_2^2+\no\\&&
((-24x-24y)z_1^2+(8+36y+36x)z_1)z_2+(-16x-16y)z_1^2+(-8x-8y)z_1)z_3^2+\no\\&&
((8y+8x)z_1+(-8x-8y)z_1^2+((-8-24x-24y)z_1^2+(36y+36x+14)z_1)z_2^2+\no\\&&
((-32x-32y)z_1^3+(-46y-46x-12)z_1-x-y+(64y+8+64x)z_1^2)z_2+x+y)z_3+\no\\&&
(4+16y+16x)z_1^2z_2^3+((16x+16y)z_1^3+(-16x-16y)z_1^2+(-8x-8y-5)z_1)z_2^2+\no\\&&
(y+x)z_1+((8y+4+8x)z_1+x+y+(-8y-8x-4)z_1^2)z_2-x-y)/\Delta,\no
\end{eqnarray}
\begin{eqnarray}
Y_{133}&=&((4+16y+16x)z_1^2z_3^3+((16x+16y)z_1^3+(-27y-9-27x)z_1z_2^2+((-8-24x-24y)z_1^2+\no\\
&&(36y+36x+14)z_1)z_2+(-16x-16y)z_1^2+(-8x-8y-5)z_1)z_3^2+((8y+4+8x)z_1+\no\\&&
(-8y-8x-4)z_1^2+((-24x-24y)z_1^2+(8+36y+36x)z_1)z_2^2+((-32x-32y)z_1^3+\no\\&&
(-46y-46x-12)z_1-x-y+(64y+8+64x)z_1^2)z_2+x+y)z_3+(16x+16y)z_1^2z_2^3+\no\\&&
((16x+16y)z_1^3+(-16x-16y)z_1^2+(-8x-8y)z_1)z_2^2+(y+x)z_1+((8y+8x)z_1+\no\\&&
x+y+(-8x-8y)z_1^2)z_2-x-y)/\Delta,
\no
\end{eqnarray}
\begin{eqnarray}
Y_{112}&=& -((48x+16y-8)z_1^2z_3^3+((48x+16y-8)z_1^3+(-27y-81x+9)z_1z_2^2+\no\\&&
((-24y-72x+14)z_1^2+(36y+108x-11)z_1)z_2+(8-16y-48x)z_1^2+\no\\&&
(-8y+2-24x)z_1)z_3^2+((8y+24x-2)z_1+(-8y+2-24x)z_1^2+((6-72x-24y)z_1^2+\no\\&&
(36y+108x-14)z_1)z_2^2+((-32y-96x+16)z_1^3+(-138x+16-46y)z_1-3x-y+\no\\&&
(64y+192x-28)z_1^2)z_2+3x+y)z_3+(48x-4+16y)z_1^2z_2^3+((48x+16y-8)z_1^3+\no\\&&
(4-48x-16y)z_1^2+(-8y+5-24x)z_1)z_2^2+(y+3x)z_1+((24x+8y-5)z_1+3x+y+\no\\&&
(-24x-8y+6)z_1^2)z_2-3x-y)/\Delta,\no\\&&
\no\\
Y_{113} 
&=&-((48x-4+16y)z_1^2z_3^3+((48x+16y-8)z_1^3+(-27y-81x+9)z_1z_2^2+\no\\&&
((6-72x-24y)z_1^2+(36y+108x-14)z_1)z_2+(4-48x-16y)z_1^2+\no\\&&
(-8y+5-24x)z_1)z_3^2+((24x+8y-5)z_1+(-24x-8y+6)z_1^2+\no\\&&
((-24y-72x+14)z_1^2+(36y+108x-11)z_1)z_2^2+((-32y-96x+16)z_1^3+\no\\&&
(-138x+16-46y)z_1-3x-y+(64y+192x-28)z_1^2)z_2+3x+y)z_3+\no\\&&
(48x+16y-8)z_1^2z_2^3+((48x+16y-8)z_1^3+(8-16y-48x)z_1^2+\no\\&&
(-8y+2-24x)z_1)z_2^2+(y+3x)z_1+((8y+24x-2)z_1+3x+y+\no\\&&
(-8y+2-24x)z_1^2)z_2-3x-y)/\Delta,
\no
\end{eqnarray}
\begin{eqnarray}
Y_{123} 
&=& ((-8+32x)z_1^2z_3^3+((-32x+8)z_1^2+(6-16x)z_1+((72x-25)z_1+\no\\&&
(14-48x)z_1^2)z_2+(-8+32x)z_1^3+(18-54x)z_1z_2^2)z_3^2+((16x-6)z_1+\no\\&&
(1+(-40+128x)z_1^2+(16-64x)z_1^3-2x+(33-92x)z_1)z_2+(6-16x)z_1^2+2x-1+\no\\&&
((72x-25)z_1+(14-48x)z_1^2)z_2^2)z_3+(-8+32x)z_1^2z_2^3+\no\\&&
((-32x+8)z_1^2+(-8+32x)z_1^3+(6-16x)z_1)z_2^2+(-1+2x)z_1+\no\\&&
((6-16x)z_1^2-1+2x+(16x-6)z_1)z_2+1-2x)/\Delta,
\no
\end{eqnarray}
\begin{eqnarray}
Y_{111}&=&((-12+96x+32y)z_1^2z_3^3+((8-32y-96x)z_1^2+(-16+96x+32y)z_1^3+\no\\&&
(-162x+18-54y)z_1z_2^2+(-48x-16y+3)z_1+((20-48y-144x)z_1^2+\no\\&&
(216x+72y-22)z_1)z_2)z_3^2+((-6x+1+(-192x+32-64y)z_1^3-2y+\no\\&&
(128y-48+384x)z_1^2+(-92y+26-276x)z_1)z_2+(8-16y-48x)z_1^2+2y+\no\\&&
((20-48y-144x)z_1^2+(216x+72y-22)z_1)z_2^2+6x+(48x+16y-2)z_1-1)z_3+\no\\&&
(-12+96x+32y)z_1^2z_2^3+((8-32y-96x)z_1^2+(-16+96x+32y)z_1^3+\no\\&&
(-48x-16y+3)z_1)z_2^2+1+(2y-2+6x)z_1+(6x+2y+(8-16y-48x)z_1^2-1+\no\\&&
(48x+16y-2)z_1)z_2-6x-2y)/\Delta.
\end{eqnarray}
\begin{eqnarray}
\Delta&=&-\bigl((-16z_2^2+32z_3z_2-16z_3^2)z_1^3+(24z_3^2z_2-16z_3^3-16z_2^3-64z_3z_2+24z_3z_2^2+16z_2^2+
\no\\&&
8z_3+16z_3^2+8z_2)z_1^2+(8z_3^2-1+27z_2^2z_3^2-36z_3z_2^2-8z_2+8z_2^2-8z_3-36z_3^2z_2+46z_3z_2)z_1+
\no\\&&
1-z_2+z_3z_2-z_3\bigr).
\end{eqnarray}

\end{document}